%% file: wileyNJDv5_AMA.tex
\documentclass[AMA,Times1COL]{WileyNJDv5} 

\articletype{Accepted at Journal of Software Evolution and Process}%

\received{Date Month Year}
\revised{Date Month Year}
\accepted{11 June 2024}
\journal{Journal of Software Evolution and Process}
\volume{00}
\copyyear{2024}
\startpage{1}

\begin{document}

\title{Semantic Similarity Loss for Neural Source Code Summarization}

\author[1]{Chia-Yi Su}

\author[1]{Collin McMillan}


\authormark{Su \textsc{et al.}}
\titlemark{Semantic Similarity Loss for Neural Source Code Summarization}

\address[1]{\orgdiv{Department of Computer Science and Engineering}, \orgname{University of Notre Dame}, \orgaddress{Notre Dame, \state{Indiana}, \country{USA}}}

\corres{Corresponding author Chia-Yi Su, Holy Cross Dr, Notre Dame, 46556, Indiana, USA. \email{csu3@nd.edu}}

\presentaddress{Holy Cross Dr, Notre Dame, 46556, Indiana, USA.}


\abstract[Abstract]{This paper presents a procedure for and evaluation of using a semantic similarity metric as a loss function for neural source code summarization. Code summarization is the task of writing natural language descriptions of source code. Neural code summarization refers to automated techniques for generating these descriptions using neural networks. Almost all current approaches involve neural networks as either standalone models or as part of a pretrained large language models e.g., GPT, Codex, LLaMA. Yet almost all also use a categorical cross-entropy (CCE) loss function for network optimization. Two problems with CCE are that 1) it computes loss over each word prediction one-at-a-time, rather than evaluating a whole sentence, and 2) it requires a perfect prediction, leaving no room for partial credit for synonyms. In this paper, we extend our previous work on semantic similarity metrics to show a procedure for using semantic similarity as a loss function to alleviate this problem, and we evaluate this procedure in several settings in both metrics-driven and human studies. In essence, we propose to use a semantic similarity metric to calculate loss over the whole output sentence prediction per training batch, rather than just loss for each word. We also propose to combine our loss with CCE for each word, which streamlines the training process compared to baselines. We evaluate our approach over several baselines and report improvement in the vast majority of conditions.}


\keywords{source code summarization, neural models, optimization, loss functions, human ratings and feedback}

\jnlcitation{\cname{%
\author{Chia-Yi Su},
\author{Collin McMillan},
}.
\ctitle{Semantic Similarity Loss for Neural Source Code Summarization} \cjournal{\it J Software: Evolution and Process} \cvol{2021;00(00):1--18}.}

\maketitle
\input introduction
\input background
\input approach
\input experiment
\input expresults
\input humanstudy

\input humanstudyresult
\input conclusion


\bmsection*{Acknowledgments}
This work is supported in part by NSF CCF-2100035 and CCF-2211428. Any opinions, findings, and conclusions expressed herein are the authors and do not necessarily reflect those of the sponsors

\bmsection*{Conflict of interest}

The authors declare no potential conflict of interests.

\bmsection*{Data Availability}

The funcom-python dataset that we propose is in our APCL Huggingface repository: \\
{\scriptsize \texttt{https://huggingface.co/datasets/apcl/funcom-python/tree/main}} \\ Also, we released the prediction files and models on LLM in our APCL Hugginface repository:\\
{\scriptsize \texttt{https://huggingface.co/apcl/funcom\_useloss/tree/main}}

\bmsection*{Code Availability}\label{sec:codeavalibility}

We release our code for experiments in our APCL Github repository, 
{\scriptsize \texttt{https://github.com/apcl-research/funcom-useloss}} 

\bibliography{wileyNJD-AMA}



\end{document}

%% file: Introduction.tex
\section{Introduction}
Source code “summaries” form the basis for programmer documentation of software. A summary is a natural language description of the behavior of a section of source code. Even a short summary such as “reads list of music files and plays them over speakers” gives a programmer an idea of the purpose of a section of code without ever needing to read the code itself. The process of writing summaries is called source code summarization \cite{haiduc2010use}. The expense of code summarization leads programmers to avoid it and drives strong research interest into automating the process. The ability to write natural language descriptions of source code on demand has long been a dream of software engineering researchers \cite{robillard2017demand}, and recent progress in neural source code summarization has drawn this dream within reach.

Neural code summarization refers to code summarization techniques based on neural networks. Almost all current related research uses neural models in one way or another (see Section \ref{sec:background}). The workhorse of most approaches is the encoder-decoder architecture in which the source code is “encoded” and a summary is “decoded” as a prediction of the model. Recent advances in academia have focused on the encoding process, such as representing the code as an abstract syntax tree (AST) \cite{leclair2019neural}, via a graph neural network (GNN) \cite{leclair2020improved}, or customized attention networks \cite{haque2020improved, tang2022ast}. In contrast, advances in industry have tended to come from scale and specialized fine-tuning. Large language models (LLMs) including GPT-4, Codex, and LLaMA have demonstrated abilities in explaining and summarizing source code. \cite{macneil2023experiences, ross2023programmer}.

Yet the training and fine-tuning procedures for practically all of these approaches are based on categorical cross-entropy loss (CCE), which does not always reflect how a human would grade model performance. This loss function is calculated at the end of each prediction and is used to update the model during training. The family of CCE calculate loss for each word as it is predicted, one word at a time. i.e., if the model has predicted “reads list of” so far and next predicts the word “sound” instead of the reference “music”, CCE will compute the loss solely on that word being incorrect. The model will be penalized solely because of that one wrong word, rather than a holistic view of “sound” being similar to “music” in the context of the sentence predicted so far. Wieting et al.~\cite{wieting2019beyond} refer to this flaw as the loss function lacking the ability to offer “partial credit.” The result is that the loss function does not reflect how humans view the output. Unlike CCE and similar functions, humans tolerate mistakes of words or grammar as long as the meaning of the sentence is unchanged.

In this paper, we evaluate the use of semantic similarity as a loss function for code summarization.  Specifically, we focus on the semantic similarity metrics between reference summary and the predicted summary. First, we extend our previous work on semantic similarity metrics~\cite{haque2022semantic} to show how to use semantic similarity as a loss function during training.  In our previous work, we introduced a metric called USE for evaluating code summaries during model testing (after training only).  Including USE as part of the training process is much more complicated, though potentially much more rewarding as work in other domains shows~\cite{wieting2019beyond}.  Yet, we do not merely duplicate previous training procedures for a software engineering problem: an additional contribution is that, unlike semantic similarity loss functions in other domains, our loss function is a drop-in replacement for CCE.  Existing baselines require an additional reinforcement learning step which adds complexity and reduces uptake by the community.  Throughout this paper, we refer to our USE-based loss function as \textbf{\texttt{use-seq}}.


We present our evaluation in two experiments. First, we train different neural code summarization techniques under standard conditions (i.e., using CCE) and with our USE-based loss function and compare them using automated metrics such as BLEU, METEOR, and USE. Second, we conduct a study with human experts who rate predicted outputs from our approach and from the model trained with the baseline CCE.

Note we conduct our experiments using several academic source code summarization models, but also an approach of our own design using fine-tuning of an industrial LLM. The advantage to using academic models is that we can control key experimental variables such as the contents of the training set. However, a disadvantage is that their small size means they tend to underperform compared to large, industrial models. So, to evaluate if our idea is still relevant at large scale, we fine-tune the LLaMa 7B model by Touvron et al.~\cite{touvron2023llama} using the LoRA procedure \cite{alpaca,alpacalora, hulora}. This fine-tuned LLM is an additional novel contribution of this paper. We compare and contrast using CCE, two other baselines, and use-seq loss. We find that our approach benefits both the compact academic models and large industrial ones.

\emph{Novelty Statement} The key novel contribution of this paper lies in: 1) our procedure for using a semantic similarity metric as a loss function instead of only an evaluation metric, and 2) our evaluation of the semantic similarity loss for the software engineering task of code summarization.  This paper is built on our own previous work of semantic similarity~\cite{haque2022semantic}, and is inspired by semantic similarity as loss in other domains~\cite{wieting2019beyond}.  However, this paper makes a novel contribution with a new procedure and evaluation for code summarization.  Our procedure is \emph{not} a duplicate of semantic similarity loss from other domains.  As Section~\ref{sec:humanstudy} will show, we make adjustments specific to the software engineering domain that are not necessarily optimal in a general purpose natural language task, yet are preferred by programmers in experiments.

\begin{table}[!b]
    \centering
	{\small
		\vspace{-0.4cm}
		\begin{tabular}{p{4cm}p{0.6cm}p{0.6cm}p{0.6cm}p{0.4cm}p{0.4cm}}
			&T          &A          &C      &   F &           \\
			Alon et al. (2019) \cite{alon2019code2seq, alon2019code2vec}	& x	&x   &   &  & \\
			LeClair et al. (2019) \cite{leclair2019neural}					& x	&x   &   & &  \\
			Nie et al. (2019) \cite{nie2019framework}					& x	&   &   &  & \\
			Haldar et al. (2020) \cite{haldar2020multi}					& x	&x   &   & &  \\
			Ahmad et al. (2020) \cite{ahmad2020transformer}				& x 	&   &   &  & \\
			Haque et al. (2020) \cite{haque2020improved}				& x	&   &x   &  &\\ 
			LeClair et al. (2020) \cite{leclair2020improved}				& x	&x &    & &\\
			Feng et al. (2020) \cite{feng2020codebert}					& x	&   &    & x & \\
			Bansal et al. (2021) \cite{bansal2021project}					& x	&   &x  & & \\ 
			Zügner et al. (2021) \cite{zugner2021languageagnostic}		& x	&   &   &   &\\
			Liu et al. (2021) \cite{liu2021retrievalaugmented}				& x	&x   &   & &  \\
   Mastropaolo et al. (2021)\cite{mastropaolo2021studying}				& x	&   &   & & x \\
   
			Kuang et al. (2022) \cite{kuang2022code}				& x	&x   &   &  & \\
			Tang et al. (2022) \cite{tang2022ast}				& x	&x   &   &   &\\
			Li et al. (2022) \cite{li2022setransformer}				& x	&x   &   & &  \\
            Khan et al. (2022) \cite{khan2022automatic}        & x &  &  & x \\
            Ahmed et al. (2022) \cite{ahmed2022few}            & x &  & x & x \\
            Gu et al. (2022) \cite{gu2022assemble}             & x &  &   & x \\
            Su et al. (2023) \cite{10.1145/3611643.3613090}    & x &  &   & x \\
            Gao et al. (2023) \cite{gao2023code}               & x & x &  &   \\
            Geng et al. (2023) \cite{geng2023interpretation}   & x &   & x &  \\
            Zhang et al. (2023) \cite{zhang2023ga}             & x & x &   &  \\
            Gao et al. (2023) \cite{gao2023encosum}            & x & x &   &  \\
            Wang et al. (2023) \cite{wang2023intra}            & x & x & x &  \\
            Geng et al. (2024) \cite{geng2024large}            & x &   & x & x \\
		\end{tabular}
	}
	\vspace{0.2cm}
	\caption{Snapshot of the past five years in source code summarization. Column $T$ means use of source code as Text. $A$ signifies learning from AST.  $C$ implies learning chiefly from code context.  $F$ means primary a fine-tuning approach for an LLM.}
	\label{tab:screlated}
\end{table}

%% file: background.tex
\vspace{-1mm}
\section{Background and Related Work}\label{sec:background}
This section discusses key related work and background.

\vspace{-3mm}
\subsection{Source Code Summarization}
Source code summarization is the task of generating short, natural language descriptions of source code. The term “code summarization” was coined by Haiduc et al.~\cite{haiduc2010use} in 2010 and the topic has been an active research area since. Between 2010 and 2017, most approaches were IR and template-based~\cite{song2019survey}. From 2017 to present, neural models have proliferated. Table~\ref{tab:screlated} shows selected papers in the last five years. The progression has been to larger models that explicitly include the code context. Different families have formed, namely AST-based such as ast-attendgru~\cite{leclair2019neural}, codegnngru~\cite{leclair2020improved}, transformer approaches~\cite{ahmad2020transformer}, and setransformer~\cite{li2022setransformer}.

Recently, LLM-based dialogue systems such as ChatGPT or tools such as Github Copilot have shown potential in describing code behavior~\cite{jin2023binary, sun2023automatic}. While these tools are not directly comparable because the underlying training data and code analysis procedures are proprietary, they are represented by LLMs fine-tuned on code summarization tasks, as we will show in the following sections. Note that our goal in this paper is to present a loss function for improving code summarization across the board, rather than claiming one single “best” neural model. Thus, our experiments cover different approaches that model code summarization as fine tuning of LLMs (see Section~\ref{sec:experiment}).  The idea of fine tuning a large language model to summarize code has been presented before~\cite{wang2021codet5, bender2021danger, 10.1145/3611643.3613090, ahmed2022few, gu2022assemble, feng2020codebert, mastropaolo2021studying} -- this paper builds on the idea with an improved procedure that trains the model with semantic similarity.

\subsection{Semantic Similarity Metrics}
Semantic similarity metrics are the automatic metrics to evaluate the source code summary. The vast majority of the automatic metrics nowadays focus on evaluating the quality between reference summary and the predicted summary such as BLEU~\cite{papineni2002bleu}, METEOR~\cite{banerjee2005meteor},CrystalBLEU~\cite{eghbali2023crystalbleU}, and USE~\cite{haque2022semantic}, which this paper relies on. Another category is to compute the similarity between source code and the summary, which benefits the case where we do not have the good reference summary. For example, Mastropaolo et al.~\cite{mastropaolo2023evaluating} proposed SIDE that gives the summary similar to the source code 1.0 and the summary dissimilar to the source code -1.0 based on contrastive learning. This paper focuses on the metrics that compare the reference summary and the predicted summary and uses the USE as an example. We leave the other metrics as our future work.

\subsection{Loss Functions in Neural Networks}
Neural models of code summarization (like models for most NLP applications), predict code summaries one word at a time. During training, the network is run several times with several inputs in a “batch.” The concept of batch is to divide the entire dataset into several small groups because we cannot fit the entire dataset into the GPU.  In most code summarization approaches, all the words for a summary for a code sample are sent in the same batch (other code samples/summaries may be in that same batch, too). Then the loss is computed over the entire batch and used to update the network.  The loss function used in, to our knowledge, all published code summarization techniques is Categorical Cross-Entropy (CCE) loss. CCE uses two values to compute loss over each word in the batch: 1) the value in the reference for that word, and 2) the network’s output at each word prediction. The value in the reference is usually a one-hot vector (denoted y) that is the length of the vocabulary size (denoted C). The network’s output is also a vocab-size-length vector (denoted $\hat{y}$), adjusted with softmax to represent the predicted probability for the word at each element location, on a 0-1 scale. The loss function formula is then:

\vspace{-2mm}
\begin{equation*}
\mathrm{CCE}(\boldsymbol{y}, \boldsymbol{\hat{y}}) = -\sum_{i=1}^{C} y_i \log{\hat{y}_i}    
\end{equation*}

Since $\boldsymbol{y}$ is usually a one-hot vector, for any given word the formula usually simplifies to $\log{\hat{y}_r}$, where $r$ is the position of the reference word.  We show hypothetical CCE values in Table~\ref{fig:useseqex} while illustrating our approach.

\vspace{-3mm}

\subsection{Semantic Similarity Loss}

Alternatives to CCE have been proposed that compute loss based on semantic similarity. Two approaches stand out as particularly relevant to code summarization: 1) using the n-gram sequence similarity metric BLEU to compute loss~\cite{ranzato2016sequence, pasunuru2018multi}, and 2) SimiLE, which uses cosine similarity of an embedding vector model to compute loss~\cite{wieting2019beyond}. SimiLE represents a family of techniques based on embedding vector similarity~\cite{chen2020deep,  nakatani2022comparing,yasui2019using}and includes improvements such as a length penalty.

We expand on more details of these approaches in Section~\ref{sec:baselines}, but the key advantage over CCE is that they calculate loss over an entire output sequence prediction instead of each individual word. However, a key disadvantage is that they require an additional reinforcement learning phase after normal CCE training that adds complexity and experimental variables. In contrast, our approach is a drop-in replacement for CCE, and our approach is designed and evaluated for the domain-specific SE task of code summarization. As we will show in Section~\ref{sec:humanstudy}, adjustments useful in general natural language such as a length penalty are not necessarily ideal for code summarization, where programmers tend to value accuracy over conciseness.

%% file: approach.tex
\section{Approach}\label{sec:approach}

This section discusses our approach to the semantic similarity loss function we propose called use-seq. The use-seq calculation is broadly divided into six steps. We show concrete outputs for each word at each step in Table~\ref{fig:useseqex}.

\begin{table}[!h]
	\small
	\centering
 \caption{Example of loss calculation for each word at each step. The batch loss is the average loss from each step.  CCE values are hypothetical for illustration.}
	\begin{tabular}{llllll}
		predicted word	& step 3	& step 4	& step 5	& (cce)			& step 6		\\ \hline
		records 		& 0.8665	& 0.8665	& 2.954		& 0.0400		& 0.1182		\\
		a 				& 0.8665	& 0.8665	& 2.954		& 0.0500		& 0.1477		\\
		sound 			& 0.8665	& m			& m			& 0.8000		& 0.8000		\\
		file 			& 0.8665	& 0.8665	& 2.954		& 0.0600		& 0.1772		\\ \hline
						&			&			& \multicolumn{2}{r}{batch loss:}& 0.3108 \\
      &			&			& \multicolumn{2}{r}{USE score:}& 0.8665
	\end{tabular}
	
	\label{fig:useseqex}
\end{table}

1) Convert predicted sequence into natural language.  During training, each training batch will consist of at least one example subroutine and summary.  Then the model will predict each position of the output one-at-a-time using the ``teacher forcing'' procedure.  For example, consider a subroutine with the reference summary ``records a music file'', but where the model makes an error by predicting the word ``sound'' instead of ``music''.  We demonstrate the prediction process for this sample in Table~\ref{tab:prediction}.

\begin{table}[!h]
        
	\centering
 \caption{ Demonstration of prediction process}
	\begin{tabular}{llll}
		& training input                                  & reference word               & predicted word\\ \hline
		1 & \textless{}s\textgreater{}                    & records                      & records \\
		2 & \textless{}s\textgreater~records              & a                            & a \\
		3 & \textless{}s\textgreater~records a            & music                        & sound \\
		4 & \textless{}s\textgreater~records a music      & file                         & file \\
		5 & \textless{}s\textgreater~records a music file & \textless{}/s\textgreater{}  & \textless{}/s\textgreater{}
	\end{tabular}
 \label{tab:prediction}
\end{table}

The model will receive the reference training input, so that the last word ``file'' is predicted using the correct previous word ``music.''  However, the total predicted sequence still contains the error.  In CCE, the loss is calculated for each predicted word to the reference word.  In our { \scriptsize \texttt{use-seq}} approach, we convert the predicted words back into a sequence, in this case ``records a sound file.'' The special tokens <s> and </s> are the start and the end tokens. The adoption of those tokens in the example is to show the entire process of the model prediction.

2) Compute semantic similarity.  The next step is to take each predicted and reference sequence, obtain the USE vector for each sequence, then compute the cosine distance between the two vectors.  The USE vector is a 512-length vector from the universal sentence encoder model~\cite{cer2018universal}.  Technically, we could use any word sequence encoder to produce these vectors, but a recent paper finds that USE produces results most in line with human preferences for code summarization~\cite{haque2022semantic}.

3) Broadcast semantic similarity to each word.  The semantic similarity calculation in {\scriptsize \texttt{use-seq}} applies to the entire sequence.  However, loss is usually calculated for each individual prediction, which in our case is each word.  We assign the loss for each word to be equal to the loss for each sequence.  For example, the cosine distance between the reference and predicted sequences above is 0.8665, so each word ``records'', ``a'', ``sound'', ``file'' will receive a loss of 0.8665.

4) Mask semantic similarities to avoid inappropriate penalties.  It is possible that USE will return a score that indicates a strongly dissimilar predicted sequence, even when some individual word predictions may be correct.  It is also possible that USE will indicate a similar sequence, even when individual word predictions may be incorrect.  For example, “disconnect db” and “initialize database connections” are two opposite sentences, but we had 0.5613 as the USE for these two sentences. Although this score is lower compared with “connect db” and “initialize database connections," which we had 0.6963 as the USE score, 0.5613 is still very high for the opposite sentence. In these situations, a naive application of sequence similarity to each word could have the effect of rewarding the model for incorrect word predictions or penalizing the model for correct word predictions.  To avoid these problems, we create a mask in which, for each word prediction, if the prediction is correct, only use the sequence similarity broadcast to that word if the cosine similarity is positive.  Likewise, if the prediction is incorrect, only use the sequence similarity broadcast if the cosine similarity is negative.   The effect is to provide an extra reward to correct predictions when the sequence similarity is high, while also giving a penalty to incorrect predictions when the sequence similarity is low.  In Table~\ref{fig:useseqex}, ``m'' denotes a mask for the word ``sound'' because ``sound'' is incorrect even though the overall sentence similarity is high as indicated in Table~\ref{fig:useseqex} that the USE score is 0.8665.

5) Adjust semantic similarities using exponentiated reward.  In preliminary experiments, we noticed that semantic similarity values tend to be distributed such that small differences from the mean seem to have less overall meaning than large ones.  In other words, semantic similarity is most useful in reporting that a sequence is very similar or dissimilar, while values around the mean are harder to interpret.  Our observation corroborates findings in using human ratings of similarity, such as by Korbak et al.~\cite{korbak2023pretraining}, and our remedy is similar: we adjust each semantic similarity score using an exponential function adjusted by parameter $\beta$.  The formula is $exp(R(x_{i})/\beta)$, where $x_{i}$ is the word prediction for position $i$ in the sequence, and $R$ is the reward function, which in our case is the USE similarity score for the sequence where $x_{i}$ originates.  The effect of this function is to push values beyond a certain threshold to have much more effect on the loss.  The value $\beta$ is a parameter which allows us to scale the similarity scores.  For example, in Table~\ref{fig:useseqex}, the scaled reward value becomes $exp(0.8665/1)=0.8665$ when $beta$ is 1. We use $\beta=1.0$ as a default value for LLMs because it is recommended by Korbak et al.~\cite{korbak2023pretraining}.  However, we explore the different values in Section~\ref{sec:ablate}. We use $\beta=0.8$ as a default value for purpose-built models because  our ablation study finds that $\beta=0.8$ shows the improvement over all datasets and across all metrics and datasets. 

6) Combine the semantic similarities with CCE for each predicted word.  A problem we noticed in preliminary experiments was that using semantic similarity scores as a loss in from-scratch training leads to very unstable and poor results -- an observation also found in using semantic similarity for other domains~\citep{yasui2019using}.  The solution in related work is to train the model to convergence using CCE, then add a fine-tuning step with semantic similarity loss.  The problem with adding a fine-tuning step is added complexity of the training procedure and creation of new experimental variables (how far to train after convergence with CCE, what parameters/methods of fine-tuning, etc.).  Our solution is instead to use semantic similarity to adjust CCE for each word (alluded to in Step 4).  We multiply the CCE for each word to the semantic similarity score for that word. We formalize entire procesure in Equation~\ref{equ:loss}.

\begin{equation*} 
R(x_{i}) =
\begin{cases}
USE &\text{  if $x_{i}$ is correct} \\
0 & \text{if $x_{i}$ if incorrect}
\end{cases}
\end{equation*}

\begin{equation}\label{equ:loss}
\mathrm{loss}(\boldsymbol{y}, \boldsymbol{\hat{y}}) = -\sum_{i=1}^{C} y_i \log{\hat{y}_i} * exp(R(x_{i})/\beta)
\end{equation}

Since $\boldsymbol{y}$ is usually a one-hot vector, for any given word the formula usually simplifies to $\log{\hat{y}_r}$, where $i$ is the position of the word; $R(x_{i})$ is the reward function; $\beta$ is the hyperparameter.  We show hypothetical CCE values in Table~\ref{fig:useseqex} while illustrating our approach.

While the description above is sufficient to reproduce our approach, given its complexity we encourage other researchers to read our implementation in the function { \texttt{custom\_use\_seq()}} at line 191 of file { \texttt{custom/qs\_loss.py}} in our reproducibility package (Section~\ref{sec:conclusion}).

%% file: experiment.tex
\section{Experiment}\label{sec:experiment}

This section describes our controlled experiment involving code summarization models and automated metrics. Note that this experiment is distinct from our human study in Section~\ref{sec:humanstudy}.
\subsection{Research Questions}
Our research objective is to test whether semantic similarity loss improves model training results for the task of source code summarization. We formalize our experiments based on Wohlin et al.~\cite{wohlin2012experimentation}. We define the independent variable as the loss function in the model and the dependent variable as the results in terms of the automatic metrics i.e. BLEU, METEOR, and USE. The hypothesis is that \texttt{use-seq} should perform better than any other loss functions in any circumstance e.g. \texttt{use-seq} should have better performance than CCE in both purpose-built models and LLMs.  Towards that end, we ask the following Research Questions (RQs):

\begin{description}
	\item[\textbf{RQ1}] What are the differences in performance among {\scriptsize \texttt{use-seq}} and the baselines for purpose-built source code summarization models?

	\item[\textbf{RQ2}] What are the differences in performance between {\scriptsize \texttt{use-seq}} and the primary baseline for select Large Language Models (LLMs) fine-tuned for source code summarization?
\end{description}

The rationale behind RQ1 is that our approach should benefit several neural network-based models of code summarization among the many that have been published in recent years.  These models tend to be relatively small (on the order of 30m-50m parameters), but are purpose-built under highly-controlled experimental settings that are available to the public.  While the performance may or may not be as high as industrial solutions such as ChatGPT or Copilot, the advantage is that we can be more certain of the effect of different experimental variables when we have access to all model details and training data.

The rationale behind RQ2 is that our approach should, in theory, benefit any model type based on neural networks, including very large ones. There are several approaches to use the pretrained Large Language Models (LLMs) for downstream tasks. We can either use prompt techniques or finetuning the models for the tasks. The advantage of those approaches is that it allows for the use of much bigger models (100m up to many billions of parameters) that tend to have better results, at least in terms of automated metrics. This paper mainly focuses on the finetuning method because our main goal is to show that the model trained with \texttt{use-seq} improves the training results.

However, the disadvantage is that the pretraining process is so expensive that it is constrained to large industrial organizations~\cite{hellendoorn2021growing}.  Many details such as the training data sets are closed source -- a major hazard for research because the training set may contain some or all of the test set~\cite{xu2022systematic}.  So, we ask RQ2 to test our approach in a fine-tuning setting to be current with the latest techniques, but we retain RQ1 as a balance in a setting where we have maximum control over experimental variables.

\subsection{Baselines} \label{sec:baselines}
There are three baselines for our work in our experiment: CCE, BLEU, and SimiLe.  Categorical Cross-entropy (CCE) loss, as we established in Section~\ref{sec:background}, is by far the main means of computing loss for text generation tasks, including code summarization.  It represents the state-of-the-practice.  BLEU is usually used as a metric for evaluation, but has been proposed as a loss function for text generation~\cite{ranzato2016sequence, pasunuru2018multi}.  Using BLEU in this fashion represents an n-gram text similarity metric.  Finally, SimiLe is a semantic text similarity metric repurposed as a loss function~\cite{wieting2019beyond}.  The technique is the most similar approach in the literature to this paper.  It combines a vector-based text similarity technique (SIM by Wieting and Gimpel~\cite{wieting2018paranmt}) with a length error penalty component.  Note that in this paper, we replaced SIM with USE as the text similarity foundation for SimiLe.  Compared with SIM, USE is newer, trained on a large corpus of text, and has a highly-supported implementation for reproducibility.  In addition, using USE reduces experimental variables as we can determine that the difference between SimiLe and other approaches is more likely to be due to the loss function formula instead of differences in the text similarity calculation.

Note that BLEU and SimiLe rely on a reinforcement learning-like training procedure that is more complex than the one needed for our approach.  As Wu et al.~\cite{wu2016google} and Yasui et al.~\cite{yasui2019using} point out, using a text similarity metric out-of-the-box as a loss function tends to lead to very unstable training results.  Therefore, the recommended procedure is to train using CCE to allow the model to converge, then fine tune using the semantic similarity loss.  In our experiments, we report BLEU and SimiLe results after one epoch of fine tuning after the CCE loss convergence.  Since the number of samples in our datasets is small (as opposed to internet-scale LLM training), we use the same learning rate and other hyperparameters during the fine-tuning epoch.

In contrast, an advantage of our approach is that we do not require the additional fine tune epoch and related procedural complexity.  Our approach is a drop-in replacement for CCE and we train models using it in the same way.

\subsection{Datasets}
We use three datasets. The first dataset, named \texttt{funcom-java}, consists of about 2 million Java methods. The dataset was originally introduced by LeClair and McMillan~\cite{leclair2019recommendations}, who advocate a split-by-project configuration to avoid data redundancy. We used the updated version of the dataset introduced by Bansal et al.~\cite{bansal2021project}, who applied additional filtering techniques to remove code clones as suggested by Allamanis~\cite{allamanis2019adverse}.

We also generated a subset of the above Java dataset, which we call \texttt{funcom-java-long} , to focus on the methods that have the higher number of code tokens and implement the key data preprocessing filters suggested by Shi et al.~\cite{shi2022are}. It contains the subroutines that have top 10\% highest number of code tokens. The reason that we focus on these functions is based on the observation made by Haque et al. ~\cite{haque2021action} that many Java functions are trivially short such as getters and setters.  The focus of the methods that have a higher number of code tokens can show that our approach is able to tackle more challenging and realistic methods because these methods are harder to understand and have less training data.

Lastly, we compiled a dataset of Python functions from 40,000 Python projects we downloaded from GitHub, named \texttt{funcom-python}. We employed the same preprocessing and splitting methods as recommended by LeClair and McMillan~\cite{leclair2019recommendations} and Bansal et al.~\cite{bansal2021project} to create a dataset of 270k functions.

\subsection{Metrics}

We use the metrics METEOR, USE, and BLEU.  BLEU by Papineni
et al.~\cite{papineni2002bleu} is an n-gram based text similarity metric used for over twenty years in several areas of research including code summarization.  METEOR by Banerjee and Lavie~\cite{banerjee2005meteor} updates the idea of BLEU to include the semantic similarity of each word.  METEOR is preferred to BLEU for code summarization in light of recent evidence finding that METEOR is more correlated to human judgment~\cite{haque2022semantic, roy2021reassessing}.  USE is the semantic similarity metric proposed by Haque et al.~\cite{haque2022semantic} that is the basis for our semantic similarity loss functions, described in Section~\ref{sec:background}.

We calculated statistical significance using a paired t-test for METEOR and USE between {\scriptsize \texttt{use-seq}} and the baselines for each code summarization technique, using the procedure suggested by Roy
et al.~\cite{roy2021reassessing} for code summarization.  However, we do not calculate statistical tests for BLEU because BLEU is a corpus-level metric and not considered reliable when calculated at sentence-level~\cite{papineni2002bleu}.  Space limitations prevent us from printing full results in this document, so we report when the test was not significant at the $p>0.05$ level with an asterisk in Table~\ref{tab:rqoneresults}.

\subsection{Code Summarization Models}
We answer our RQs in the context of four purpose-built code summarization models.  These models represent different families of models that we identified in Section~\ref{sec:background}.  In our view, these approaches serve as ``mouse models'' which may have some distance from in-practice use, but have the major advantage that we can tightly regulate experimental variables including complete training in the laboratory setting and reproducibility at reasonable cost.  The four models are:

\textbf{ast-attendgru} An approach that encodes the target function's Abstract Syntax Tree (AST) via a flattened representation and a recurrent neural network (a GRU)~\cite{leclair2019neural}.

\textbf{codegnngru} An approach by LeClair et al. ~\cite{leclair2020improved} that is similar to {\texttt{ast-attendgru}} except that it uses a graph neural network (GNN) to encode the AST.

\textbf{transformer} Essentially this approach uses a vanilla Transformer encoder-decoder design, proposed for use on code summarization by Ahmad et al.~\cite{ahmad2020transformer}.

\textbf{setransformer} A hybrid Transformer-CNN model proposed recently by Li et al.~\cite{li2022setransformer} that encodes the AST and textual information from the code.

We train all four models from scratch on the training set from each dataset.  Our main interest is to test our loss function rather than other variables, so we follow the training procedure established by several recent papers~\cite{bansal2021project, haque2020improved, leclair2020improved}: train for ten epochs, select the epoch for which the validation accuracy was the highest, then report metric scores over the testing set for that epoch.  Key hyperparameters are shown in Table~\ref{tab:hyperparameter}:

\begin{table}[h!]
	\centering
	\small
 \caption{Training hyperparameters and settings}
	\begin{tabular}{llll}
		&                                       & Java & Python \\
		$t$ & tokens in target subroutine & 50 & 50 \\
		$w$ & words in summary            & 13 & 13 \\
		$v$ & source code vocabulary size~~           & 75k & 100k \\
		$z$ & summary vocabulary size               & 10908 & 11000 \\
		$e$ & embedding dimensions                  & 100 & 100 \\
		$b$ & batch size                            & 50 & 50 \\
		$r$ & learning rate                         & 1e-4 & 1e-4 \\   
		$o$ & optimizer								& Adam & Adam \\
	\end{tabular}
 \label{tab:hyperparameter}
\end{table}

We used $t$ and $w$ reported by Haque et al.~\cite{haque2020improved} and Bansal et al.~\cite{bansal2021project}.  The values for $v$ and $z$ are suggestions from a study of code summarization datasets~\cite{leclair2019recommendations}.  We decided $e$, $b$, $r$, and $o$ during pilot studies and constrained by hardware limitations -- our goal is for our experiments to be reproducible with moderately-priced professional hardware.

\subsection{Fine-tuned Large Language Models}
\label{sec:finetuned}

We also answer our RQs in the context of fine-tuned large language models (LLMs).  State-of-the-art performance in many text generation tasks is often produced by fine-tuning so-called foundation models.  A foundation model is an LLM that is pre-trained on internet-scale text datasets.  Then the foundation model is fine-tuned by further training on a relatively small dataset of domain-specific examples.  There are hundreds of possible fine-tuning configurations, and a comprehensive study of fine-tuning for source code summarization is not available yet in the literature and is beyond the scope of this single paper.  However, our goal is to determine the usefulness of our semantic similarity loss function in a wide range of models, so we chose two approaches consistent with related work for other text generation tasks.  Given the high cost of fine-tuning LLMs and the immense number of experimental variables given the rapidly changing research frontier, we limit ourselves to comparing { \scriptsize\texttt{use-seq}} to CCE for the { \texttt{funcom-java-long}} dataset.

One LLM we use is the LLaMA 7B parameter model by Touvron et al.~\cite{touvron2023llama}.  We fine-tune this model with the Alpaca-LoRA procedure using the settings and implementation available from Taori et al.~\cite{alpaca}.  Technically, we set the { \texttt{Instruction}} text to ``please describe the following source code'', the { \texttt{Input}} text to the target function's source code, and the { \texttt{Response}} text to the source code summary during fine tuning.  Then we fine tune for one epoch using the default parameters (listed in our reproducibility package, Section~\ref{sec:conclusion}).  During inference, we use the same { \texttt{Instruction}} and { \texttt{Input}} text setup, but extract the text after { \texttt{Response}} as the summary.

We also use the GPT2 124m parameter model by Radford et al.~\cite{radford2019language}.  We use a complete fine-tune procedure (as opposed to a weight matrix reduction technique such as LoRA) based on OpenAI's GPT2 124m parameter snapshot.  We use hyperparameters recommended by Karpathy~\cite{nanogpt} in a public GPT2 implementation.  We fine tune for 18 epochs, at which point validation accuracy diverges from training accuracy, indicating possible overfitting.  We acknowledge that these parameter choices are somewhat arbitrary, but our objective is to test the effect of training with CCE versus {\scriptsize \texttt{use-seq}}, rather than an exhaustive search for all optimal parameters.  Our idea was to use reasonable settings from related work, and compare performance differences when changing only a single experimental variable: the loss function.

In addition to 124m parameter GPT2, we use 220m CodeT5+~\cite{wang2021codet5} with complete finetuning procedure to show that our \texttt{use-seq} loss can be applied to broader language models i.e. the large language model with the encoder-decoder architecture. We finetune for three epochs. We use the hyperparameters set in the finetuning script in the CodeT5+ repository. We use the same hyperparameters for both CCE and \texttt{use-seq} because our main goal is to show that the model trained with \texttt{use-seq} is better than the model trained with CCE loss.

Note that we intend for these models to be representative of applying the technique to fine-tuning commercial LLMs, to the extent possible in a research setting where we exert control of experimental variables and are able to open source release experimental artifacts.  As Hellendoorn and Sawant~\cite{hellendoorn2021growing} point out, large, closed source LLMs such as GPT3.5 or GPT4 produce strong results at the expense of accessibility of the internals to the research community.  Smaller models capable of being run in-house also have the major advantage of not requiring proprietary code data to be sent to a third party.  The loss of data custody required to use e.g. GPT4 is not tenable for many organizations.  Furthermore, emerging evidence is showing that commercial LLMs do not always provide stable results, making them difficult to benchmark in a controlled experiment~\cite{jin2023binary, sun2023automatic}.  Therefore, we evaluate our idea in the situation of in-house, open source production.

\subsection{Ablation Study}
\label{sec:ablate}
To show the necessity of the reward mechanism that we introduce in Section~\ref{sec:approach} and study the different $\beta$ value, we conduct the ablation study with the GPT2 model. We used the GPT2 model because this model does not require weight matrix reduction technique to finetune. Also, this model has the best performance among the models that do not need weight matrix reduction technique. As an additional verification, we also used the dataset proposed by Su and McMillan~\cite{su2024distilled} to finetune the GPT2 model with the parameters that we introduce in Section~\ref{sec:finetuned}. Specifically, we focus on 170k dataset because it has exactly the same method as in \texttt{funcom-java-long}. However, instead of obtaining summary from human programmers, the summary of this dataset is generated by using GPT-3.5.

\subsection{Hardware/Software Details}

Our hardware platform is a workstation with an AMD 5900X CPU, 128GB memory, and two Nvidia A5000 GPUs.  Our software platform consists of Ubuntu 22.04, Python 3.10, CUDA 11.4, Tensorflow 2.9.1, and Pytorch 2.0.0.

%% file: expresults.tex
\section{Experimental Results}
This section discusses our experimental results and answers to research questions RQ1 and RQ2.

\subsection{RQ1: Purpose-built Models}

Table~\ref{tab:rqoneresults} summarizes our experimental results for RQ1.  In short, we find that { \texttt{use-seq}} outperforms the baselines in most conditions over the three datasets and four purpose-built code summarization models.  We observe the strongest overall performance in the { \texttt{transformer}} model, with 34.16 METEOR and 52.23 USE scores in the { \texttt{funcom-java-long}} dataset.  These are about a 2\% and 3\% improvement over SimiLE and CCE for these metrics in { \texttt{funcom-java-long}}.  In { \texttt{funcom-java}} and { \texttt{funcom-python}}, the improvements are mostly in the 1-2\% range.

It is important to note that these improvements come at practically no cost, since {  \scriptsize\texttt{use-seq}} is a drop-in replacement for CCE.  These improvements cover a broad spectrum, as we observe them over datasets with two different programming languages and over several model architectures.  So, even a modest increase in metrics scores can have a big impact on the state-of-the-art because the increases benefit many approaches at almost no cost.

We observe higher rates of improvement using BLEU score, with increases in the 2-7\% range for { \texttt{transformer}} over the three datasets.  The highest performing model in terms of BLEU is { \texttt{ast-attendgru}} with { \scriptsize \texttt{use-seq}} over { \texttt{funcom-python}}, which at 20.12 BLEU is 12\% higher than the same model using SimiLE and 18\% higher than CCE.  We attribute this difference between BLEU and METEOR/USE to BLEU's use of exact-match n-grams versus METEOR/USE's word similarity approach.  Our approach includes a mask that scales the reward based on sentence similarity, but retains a high penalty for incorrect words even in ``good'' sequence predictions (Section~\ref{sec:approach}, step 4).  In contrast, the BLEU and SimiLE functions reward the model when the overall sequence is ``good'', even when individual word predictions are incorrect.  The result is that our approach gets more individual word predictions correct, and this result manifests itself as bigger gains for BLEU than METEOR or USE.

Two exceptions to the general rule of improvement are: 1) METEOR/USE scores for { \texttt{setransformer}} over { \texttt{funcom-python}}, and 2) a handful of results that lack statistical significance, especially for { \texttt{codegnngru}} over { \texttt{funcom-python}}.  One explanation is that { \texttt{setransformer}} and { \texttt{codegnngru}} rely more than other models on the code's AST (even { \texttt{ast-attendgru}}, which has separate encoders), which is often less informative in Python (e.g., due to dynamic typing), which was also observed by Tang et al.~\cite{tang2022ast}.

\input rqoneresults

\subsection{RQ2: Large Language Models}

We find that {  \scriptsize\texttt{use-seq}} improves fine-tuning results by a statistically-significant margin for all of the LLM models i.e. GPT2, LLaMA, and CodeT5+ models.  Table~\ref{tab:rqtworesults} shows these results for RQ2.  For GPT2, we observe a 8\% improvement in METEOR, a 4\% improvement in USE, and a 8\% improvement in BLEU.  For LLaMA, we observed improvements of 3\% and 5\% for METEOR and BLEU. USE was a statistical tie with only a 0.2\% increase that was not found to be significant. Although CodeT5+ has the least improvement among these models, we still observe at least 1\% improvement with a statistically-significant margin in both METEOR and USE.   These results are in broad agreement with findings from RQ1.

These results point to the usefulness of { \scriptsize \texttt{use-seq}} even at scale.  The purpose-built code summarization models that we used are in the range of 30-50m parameters \cite{leclair2020improved, ahmad2020transformer, li2022setransformer}, while the GPT2 model we used is 124m parameters, the CodeT5+ is 220m, and the LLaMA model we used is 7B parameters.  The purpose-built code summarization models have no pretraining data, and the GPT2, CodeT5+, and LLaMA models each have different sets of pretraining data composed of differing amounts of text, code, and other types of language artifacts.  It is likely that additional hyperparameter tuning and model optimization would yield higher overall scores, though the evidence here is that {  \scriptsize\texttt{use-seq}} confers an advantage to a wide variety of models under different conditions.  Meanwhile, baselines except CCE are not practical for these larger models as they would require at least one additional training epoch (around 60 hours for { \texttt{funcom-java-long}} for LLaMA on our hardware).

\input rqtworesults

\subsection{Results of Ablation Study}
Table~\ref{tab:ablaterewardresults} summarizes the results of the ablation study for reward mechanism. Overall, the model with the reward mechanism that we introduce shows the strongest improvement across three metrics and two different datasets. Specifically, we find  a 5\% improvement in  METEOR, a 4\% improvement in USE, and an 8\% improvement in BLEU with the reward mechanism on \texttt{funcom-java-long} dataset compared with the model without reward mechanism. For GPT-3.5,  we find the  5\% improvement in METEOR, 2\% improvement in USE, and the 14\% improvement in BLEU. Moreover, we find 11\% drop in BLEU score when we compare CCE with the model without the reward mechanism on GPT-3.5 summary. These results show the necessity of the reward mechanism to adjust  the final reward, so the models do not reward the incorrect prediction and penalize the correct prediction.

Table~\ref{tab:ablatebetasults} shows the results for different $\beta$ values. In short, we find $\beta=0.8$ has the improvements over three different metrics i.e. BLEU, METEOR, and USE and two different datasets compared with $\beta=1.0$. Specifically, we observe that $\beta=0.8$ has the strongest improvement on USE and BLEU, which has a 0.7\% improvement on USE and a 2\% improvement on BLEU among these different $\beta$ values on the \texttt{funcom-java-long} dataset. Although  $\beta=0.8$ does not have the strongest improvement in terms of GPT-3.5 summary, the improvement between $\beta=0.8$ and $\beta=0.6$ is  relatively small compared with the improvement between $\beta=0.8$ and $\beta=1.0$. The improvement between $\beta=0.8$ and $\beta=1.0$ is 1\% versus 0.1\% improvement between $\beta=0.8$ and $\beta=0.6$ in terms of METEOR. 
These results show that $\beta=0.8$ is an appropriate choice because it shows improvements over a wide variety of datasets and metrics.

\input ablationrewardresult

\input ablationbetaresult

\subsection{Examples}
\label{sec:experimentresults}
We provide four examples in Table~\ref{tab:examples} to illustrate different scenarios of how the models perform.  Example 1 shows a method called { \texttt{draw()}} where { \scriptsize \texttt{use-seq}} resulted in better summaries throughout the experiment.  The fine-tuned LLaMA model predicted an addition modifier ``control'' with { \scriptsize \texttt{use-seq}}, which not only more-closely matched the reference, but provided an extra word that humans reported as being more accurate (see our human study in the next section).  Likewise, Example 2 shows how {  \scriptsize\texttt{use-seq}} led the LLaMA model to output additional relevant information compared to CCE.  In general, the predictions using { \scriptsize \texttt{use-seq}} are longer than with CCE, likely because during training, predictions of the end-of-sequence token will be penalized especially heavily when they are both wrong and cause the sequence to lack information from the reference.  If the predicted sequence is not similar enough to the reference, the USE score will be low (Section~\ref{sec:approach}, step 3), and that low score will amplify the penalty of a mispredicted token such as the end of sequence token (Section~\ref{sec:approach}, step 4). 

Examples 3 and 4 show where {  \scriptsize\texttt{use-seq}} does not necessarily always help.  Example~3 is an oddity because it is an exception to the general rule of more verbose summaries from { \scriptsize \texttt{use-seq}}, as it missed the word ``xy'' modifying dataset.  It also shows how human sometimes prefer the more verbose summaries, even if the metrics scores are lower.  Example~4 shows a similar situation for LLaMA, where the more verbose summary gets lower metrics scores, but actually includes more relevant information in the summary.  These situations tend to favor {  \scriptsize\texttt{use-seq}} because { \scriptsize \texttt{use-seq}} generally leads to more verbose predictions.  We will explore the issue of verbosity in summaries more in our human study in the next section.

\input exone

\subsection{Threats to Validity}

We divide Key threats to validity into three different categories i.e. internal, construct, and external threats to validity based on Wohlin et al.~\cite{wohlin2012experimentation}. The internal threats to validity include datasets and the pretrainng data used in the LLMs. The datasets are a threat to validity because different test inputs could yield very different results.  To help mitigate this risk, we use datasets over two languages (Java and Python), with a special emphasis on subroutines with more code tokens in Java.  key risk to the LLMs is that the training data is closed source and we cannot guarantee that the test sets (which are derived from open source projects) are not in the training set. We aim to mitigate this risk by contrasting the LLM part of the experiment from the code summarization models (for which there is no pretraining data) and by using LLMs that are reportedly trained on different datasets. The construct threats to validity is the code summarization models and LLMs we train/fine-tune. The code summarization models, training, and fine-tuning procedures can also affect the results, as e.g., more epochs for training may yield better or worse results.  We mitigate this risk by using established experimental procedures in code summarization models and by making diverse choices for the LLMs (e.g., LoRA versus a complete fine-tuning, LLaMA versus GPT2 and CodeT5+) to decrease the risk of the results being meaningful in only a special setting. The external threats to validity is \texttt{use-seq} might not be able to be applied to the most up-to-date LLMs. We mitigate this by conducting the experiments with three different types of LLMs i.e. LLaMA, GPT2, and CodeT5+. 

%% file: rqoneresults.tex
\begin{table*}[!h]
	\centering
	\small
	\caption{Results of automatic metrics for RQ1.  M=METEOR, U=USE, B=BLEU.  W is the number of times \texttt{use-seq} was the highest over all metrics and datasets.  An asterisk indicates METEOR or USE results that are not statistically different from \texttt{use-seq} according to a paired t-test at the $p<0.05$ level.}
	\begin{tabular}{lllllllllll||l}
		&                               & \multicolumn{3}{c}{funcom-java-long}                                       & \multicolumn{3}{c}{funcom-java}                                            & \multicolumn{3}{c}{funcom-python}  &                                    \\
		model         & loss                          & \multicolumn{1}{c}{M} & \multicolumn{1}{c}{U} & \multicolumn{1}{c}{B}      & \multicolumn{1}{c}{M} & \multicolumn{1}{c}{U} & \multicolumn{1}{c}{B}      & \multicolumn{1}{c}{M} & \multicolumn{1}{c}{U} & \multicolumn{1}{c}{B} & W \\ \hline
		ast-attendgru & \multicolumn{1}{l|}{cce}      &    33.21              &   50.12             & \multicolumn{1}{l|}{18.94} &        35.30              &        52.89               & \multicolumn{1}{l|}{18.33}      &        26.48               &        43.27               &       16.96       &         \\
		& \multicolumn{1}{l|}{bleu}     &  33.43                     &            50.02           & \multicolumn{1}{l|}{18.92}      &          35.56             &            53.03           & \multicolumn{1}{l|}{18.68}      &         26.41             &      42.37                 &      17.64              &   \\
		& \multicolumn{1}{l|}{simile}   &      33.34            &         49.87       & \multicolumn{1}{l|}{18.87} &  35.68                  &        53.18               & \multicolumn{1}{l|}{18.77}      &             26.54         &         42.66              &      17.89           &      \\
		& \multicolumn{1}{l|}{use-seq}  &       \textbf{33.74}        &     \textbf{50.52}                & \multicolumn{1}{l|}{ \textbf{19.38}} &      \textbf{35.96}                 &         \textbf{53.74}              & \multicolumn{1}{l|}{\textbf{19.07}}      &     \textbf{28.42}                  &              \textbf{44.00}         &      \textbf{20.12}         & 9/9        \\ \hline
		codegnngru    & \multicolumn{1}{l|}{cce}      &  32.98                     &     49.85                  & \multicolumn{1}{l|}{18.75}      &     35.82                  &            53.26           & \multicolumn{1}{l|}{18.77}      &       25.32                &              41.86         &            16.80    &       \\
		& \multicolumn{1}{l|}{bleu}     &    32.53                   &           49.46            & \multicolumn{1}{l|}{18.66}      &       35.64                &          53.21             & \multicolumn{1}{l|}{18.61}      &          25.74*             &           42.14*            &           \textbf{16.93}   &         \\
		& \multicolumn{1}{l|}{simile}   &       32.47              &       49.52               & \multicolumn{1}{l|}{18.66}      &         36.03*             &         53.43              & \multicolumn{1}{l|}{18.66}      &           25.56*          &        42.13*               &    16.81        &           \\
		& \multicolumn{1}{l|}{use-seq}  &         \textbf{33.97}              &          \textbf{50.92}            & \multicolumn{1}{l|}{\textbf{19.51}}      &        \textbf{36.17}            &           \textbf{53.84}            & \multicolumn{1}{l|}{\textbf{19.25}}      &           \textbf{25.89}           &       \textbf{42.17}               &           16.51    & 8/9        \\ \hline
		transformer   & \multicolumn{1}{l|}{cce}      &   33.57                    &    51.60                   & \multicolumn{1}{l|}{19.07} &         35.86         &    53.89             & \multicolumn{1}{l|}{18.39} &       26.97           &      44.02           &       15.70     &     \\
		& \multicolumn{1}{l|}{bleu}     &      33.52                 &          51.60             & \multicolumn{1}{l|}{18.92}      &     35.92                  &      53.84                 & \multicolumn{1}{l|}{18.60}      &        26.89               &             43.62          &         16.37       &       \\
		& \multicolumn{1}{l|}{simile}   &      33.40                 &     51.31                  & \multicolumn{1}{l|}{18.90} &    35.98                   &       53.79               & \multicolumn{1}{l|}{18.63}      &    26.87                   &             43.76        &        16.35       &      \\
		& \multicolumn{1}{l|}{use-seq}  &         \textbf{34.16}              &           \textbf{52.23}            & \multicolumn{1}{l|}{\textbf{19.63}} &       \textbf{36.17}                &       \textbf{54.45}                & \multicolumn{1}{l|}{\textbf{18.97}}      &            \textbf{27.38}           &        \textbf{44.31}               &          \textbf{17.56}     & 9/9        \\ \hline
		setransformer & \multicolumn{1}{l|}{cce}      &   32.60                    &     49.56                  & \multicolumn{1}{l|}{18.38}      &       35.64                &         53.09              & \multicolumn{1}{l|}{18.26}      &        27.92               &            43.48           &           18.10      &      \\
		& \multicolumn{1}{l|}{bleu}     &     32.21                  &         48.80              & \multicolumn{1}{l|}{18.49}      &      35.99                 &         53.36              & \multicolumn{1}{l|}{18.69}      &       28.59*                &      43.94                 &           18.76     &       \\
		& \multicolumn{1}{l|}{simile}   &   32.22                 &          48.66          & \multicolumn{1}{l|}{18.48}      &     36.03*                 &        53.43             & \multicolumn{1}{l|}{18.66}      &       \textbf{28.61*}                &              \textbf{44.05}         &              18.91   &      \\
		& \multicolumn{1}{l|}{use-seq}  &       \textbf{33.51\textcolor{white}{*}}                &           \textbf{50.79\textcolor{white}{*}}            & \multicolumn{1}{l|}{\textbf{19.04\textcolor{white}{*}}}      &    \textbf{36.35\textcolor{white}{*}}                  &       \textbf{54.14\textcolor{white}{*}}                & \multicolumn{1}{l|}{\textbf{19.12\textcolor{white}{*}}}      &        28.57\textcolor{white}{*}               &             44.02\textcolor{white}{*}          &   \textbf{19.46\textcolor{white}{*}}         & 7/9      
	\end{tabular}
\vspace{3mm}

	\label{tab:rqoneresults}
\end{table*}

%% file: rqtworesults.tex
\begin{table}[!h]
	\small
	\centering
 \captionsetup{justification=centering}
	\caption{Results of automatic metrics for RQ2. An asterisk indicates METEOR or USE results that are not statistically
different from use-seq according to a paired t-test at the p < 0.05 level.}
	\begin{tabular}{lllll}
		&                              & \multicolumn{3}{c}{funcom-java-long}                                       \\
		model   & loss                         & \multicolumn{1}{c}{M} & \multicolumn{1}{c}{U} & \multicolumn{1}{c}{B}      \\
		gpt2    & \multicolumn{1}{l|}{cce}     &                32.77  &          51.37        & \multicolumn{1}{l}{19.02} \\
		& \multicolumn{1}{l|}{use-seq} & \textbf{35.51}                 & \textbf{53.50}                 & \multicolumn{1}{l}{\textbf{20.58}} \\ \hline
		llama7b & \multicolumn{1}{l|}{cce}     & 39.60\textcolor{white}{*}                  & 59.60*                  & \multicolumn{1}{l}{23.22\textcolor{white}{*}} \\
		& \multicolumn{1}{l|}{use-seq} & \textbf{40.87}                 & \textbf{59.73}                 & \multicolumn{1}{l}{\textbf{24.40}}  \\ \hline 
  codet5+ & \multicolumn{1}{l|}{cce}     & 17.33\textcolor{white}{*}                  &      46.58      & \multicolumn{1}{l} {\textbf{0.23\textcolor{white}{*}}} \\
		& \multicolumn{1}{l|}{use-seq} & \textbf{17.53}                 & \textbf{47.20}                 & \multicolumn{1}{l}{0.00}  \\ 
	\end{tabular}
    
	\label{tab:rqtworesults}
    \vspace{-1mm}
\end{table}

%% file: ablationrewardresult.tex
\begin{table}[h]
	\small
	\centering
 \captionsetup{justification=centering}
	\caption{Result of ablation study on reward mechanism. cce means the model trained with the cce loss. use-seq means the model trained with the use-seq loss with the reward mechanism that we proposed. use-seq-ablate means the usq-seq loss without reward mechanism that we proposed.}
	\begin{tabular}{lllll}
	
		dataset   & loss                         & \multicolumn{1}{c}{M} & \multicolumn{1}{c}{U} & \multicolumn{1}{c}{B}      \\
		funcom-java-long    & \multicolumn{1}{l|}{cce} &              32.77    &    51.37        & \multicolumn{1}{l}{19.02 } \\
  & \multicolumn{1}{l|}{use-seq-ablate} &       33.76           &       51.48       & \multicolumn{1}{l}{ 19.12} \\
        & \multicolumn{1}{l|}{use-seq} & \textbf{35.51}                 & \textbf{53.50}                 & \multicolumn{1}{l}{\textbf{20.58}} \\
  \hline
		GPT-3.5 summary 	& \multicolumn{1}{l|}{cce} &  33.68             &     62.42         & \multicolumn{1}{l}{12.90}  \\   
  & \multicolumn{1}{l|}{use-seq-ablate} &     37.43          &  63.30            & \multicolumn{1}{l}{11.42}  \\   
  & \multicolumn{1}{l|}{use-seq} & \textbf{39.35}                 &\textbf{64.78}                 & \multicolumn{1}{l}{\textbf{13.00}}\\
	\end{tabular}
    
	\label{tab:ablaterewardresults}
    \vspace{-1mm}
\end{table}

%% file: ablationbetaresult.tex
\begin{table}[h]
	\small
	\centering
 \captionsetup{justification=centering}
	\caption{Result of ablation study on selection of $\beta$}
	\begin{tabular}{lllll}
	
		dataset   & $\beta$                         & \multicolumn{1}{c}{M} & \multicolumn{1}{c}{U} & \multicolumn{1}{c}{B}      \\
		funcom-java-long    & \multicolumn{1}{l|}{1.2} &         35.71          &    51.99       & \multicolumn{1}{l}{ 17.22  } \\
  & \multicolumn{1}{l|}{1.0} &      35.51             &     53.50         & \multicolumn{1}{l}{ 20.58 } \\
    & \multicolumn{1}{l|}{0.8} &      35.68             &      \textbf{53.88}         & \multicolumn{1}{l}{ \textbf{20.94} } \\
        & \multicolumn{1}{l|}{0.6} & \textbf{36.05}                 & 52.29                & \multicolumn{1}{l}{17.14} \\
  \hline
		GPT-3.5 summary 	& \multicolumn{1}{l|}{1.2} &   39.63            &       65.00       & \multicolumn{1}{l}{13.49}  \\   
  & \multicolumn{1}{l|}{1.0} &        39.35       &   64.78           & \multicolumn{1}{l}{13.00}  \\  
  & \multicolumn{1}{l|}{0.8} &       39.83        &        65.01      & \multicolumn{1}{l}{13.36}  \\  
  & \multicolumn{1}{l|}{0.6} & \textbf{39.90}                 &\textbf{65.31}                 & \multicolumn{1}{l}{\textbf{13.70}}\\
	\end{tabular}
    
	\label{tab:ablatebetasults}
    \vspace{-1mm}
\end{table}

%% file: exone.tex
\begin{table*}[!h]
\caption{The prediction examples.  Transformer means of output from the purpose-built model.  llama7b refers to the output from finetuned Llama.  M=METEOR, U=USE.  H=1 means summary was considered more accurate in the human study (dash means summary not shown in human study).  The number underneath the method name is the identification number of that method in the \texttt{funcom-java-long} dataset, which we provide for reproducibility.  Note that for illustration we deliberately chose some examples where \texttt{use-seq} does not show improvement.}
    \vspace{2mm}
	\begin{adjustwidth}{-0.2cm}{}
	{\setlength{\tabcolsep}{0.5em}
	\centering
	\small
	\begin{tabular}{p{0.0cm}p{2.4cm}p{1.2cm}p{0.95cm}p{9.3cm}|lll}
		1. &  draw()	   & reference   &         & draw the control fps panel in the control sketch window & M & U & H   \\ \cline{3-5}
	& {\scriptsize \emph{7848915}}& transformer & cce     & draws the graphic & 10.75 & 26.12 & -  \\
	& &             & use-seq & draws the outline of the current page & \textbf{26.35} & \textbf{27.36} & - \\ \cline{6-8}
	& & llama7b       & cce     & draw the fps & 16.13 & 37.20 & 0 \\
		& &             & use-seq & draws the control fps & \textbf{33.58} & \textbf{39.54} & 1  
	\end{tabular}

	\vspace{3mm}

	\begin{tabular}{p{0.0cm}p{2.4cm}p{1.2cm}p{0.95cm}p{9.3cm}|lll}
	2. & destroyMainPart() & reference   &         & destroys one of the main parts of the given docking graph & M & U & H   \\ \cline{3-5}
	& {\scriptsize \emph{13278733}} & transformer & cce     & removes the main part of the graph & 55.56 & 75.73 & -   \\
	& &             & use-seq & destroys the main part of the graph & \textbf{63.44} & \textbf{81.56} & -  \\ \cline{6-8}
	& & llama7b       & cce     & destroys the main part & 30.64 & 56.77 & -\\
	& &             & use-seq & destroy the main part of the docking graph & \textbf{72.79} & \textbf{82.85} & -
	\end{tabular}
	
	\vspace{3mm}
	
	\begin{tabular}{p{0.0cm}p{2.4cm}p{1.2cm}p{0.95cm}p{9.3cm}|lll}
	3. & createDataset() & reference   &         & creates a sample dataset & M & U & H  \\ \cline{3-5}
	& {\scriptsize \emph{35061399}}& transformer & cce     & creates a dataset for the chart & \textbf{10.64} & 10.55 & -  \\
	& &             & use-seq & creates a dataset  & 10.00 & \textbf{36.93} & - \\ \cline{6-8}
	& & llama7b       & cce     & creates an xy dataset & 10.64 & 10.55 & 1\\
	& &             & use-seq & creates a dataset & \textbf{19.94} & \textbf{17.68} & 0
	\end{tabular}

	\vspace{3mm}
	\begin{tabular}{p{0.0cm}p{2.4cm}p{1.2cm}p{0.95cm}p{9.3cm}|lll}
	4. & copySubstring() & reference   &         & copy string to clipboard & M & U & H  \\ \cline{3-5}
	& {\scriptsize \emph{40467691}}& transformer & cce     & copies the substring of the specified & 11.90 & 50.24 & -  \\
	& &             & use-seq & copies the substring of the code string code   & \textbf{22.72} & \textbf{57.02} & -\\ \cline{6-8}
	& & llama7b       & cce     & copies a substring of the text to the clipboard & \textbf{33.33} & \textbf{69.99} & -\\
	& &             & use-seq & copies the substring from the given start to the given end to the clipboard & 30.00 & 47.25 & -
	\end{tabular}

	}
	\end{adjustwidth}

    \vspace{3mm}
	
	\label{tab:examples}
	
	\vspace{-1mm}

\end{table*}

%% file: humanstudy.tex
\section{Human Study}\label{sec:humanstudy}

This section describes our study with human programmers.  In short, we recruited human experts to compare the summaries generated by the LLaMA 7B model that we fine-tuned using CCE and {  \scriptsize\texttt{use-seq}}.

\subsection{Research Questions}

The objective of the human study is to show that \texttt{use-seq} not only has better performance in terms of automatic metrics but is preferred by the programmers.  The independent variable for this experiment is loss functions. The dependent variable is the results from the human experts. The hypothesis is that the improved summary should reflect on the results of both automatic metrics and the human experts. Therefore, we ask the following research questions:

\begin{description}
	\item[\textbf{RQ3}] How do human experts rank { \texttt{cce}} and {  \scriptsize\texttt{use-seq}} in terms of the quality attributes accuracy, completeness, conciseness, and similarity to a reference?
	\item[\textbf{RQ4}] Which of {\texttt{cce}} and { \scriptsize\texttt{use-seq}} do human experts prefer overall, independent of individual quality attributes?
\end{description}

The rationale behind RQ3 is that many years of studies in source code summarization use four quality attributes to compare summaries.  These are: 1) Accuracy, which refers to whether the information in the summary is correct, 2) Completeness, which refers to whether the summary contains all the information it should, 3) Conciseness, which refers to whether the summary is overly verbose or contains unnecessary information, and 4) Similarity to a reference, which is a human judgment about whether a summary is similar to the summary for a code in the gold set.  These quality attributes have a very long history in evaluating code comments~\cite{sridhara2010towards, ferretti2023naturalness, rani2023decade}, so we use them as the basis for evaluating our approach.

The rationale behind RQ4 is that human experts may have their own subjective opinion about the quality difference between two summaries, and this opinion may be separate from the quality attributes typically studied in the literature.  People are unique, and sometimes may prefer one thing over another for unpredictable reasons.  We study RQ4 to capture these unpredictable preferences.

\subsection{Study Design}

\begin{figure}[t]
	\centering
	\includegraphics[width=17cm]{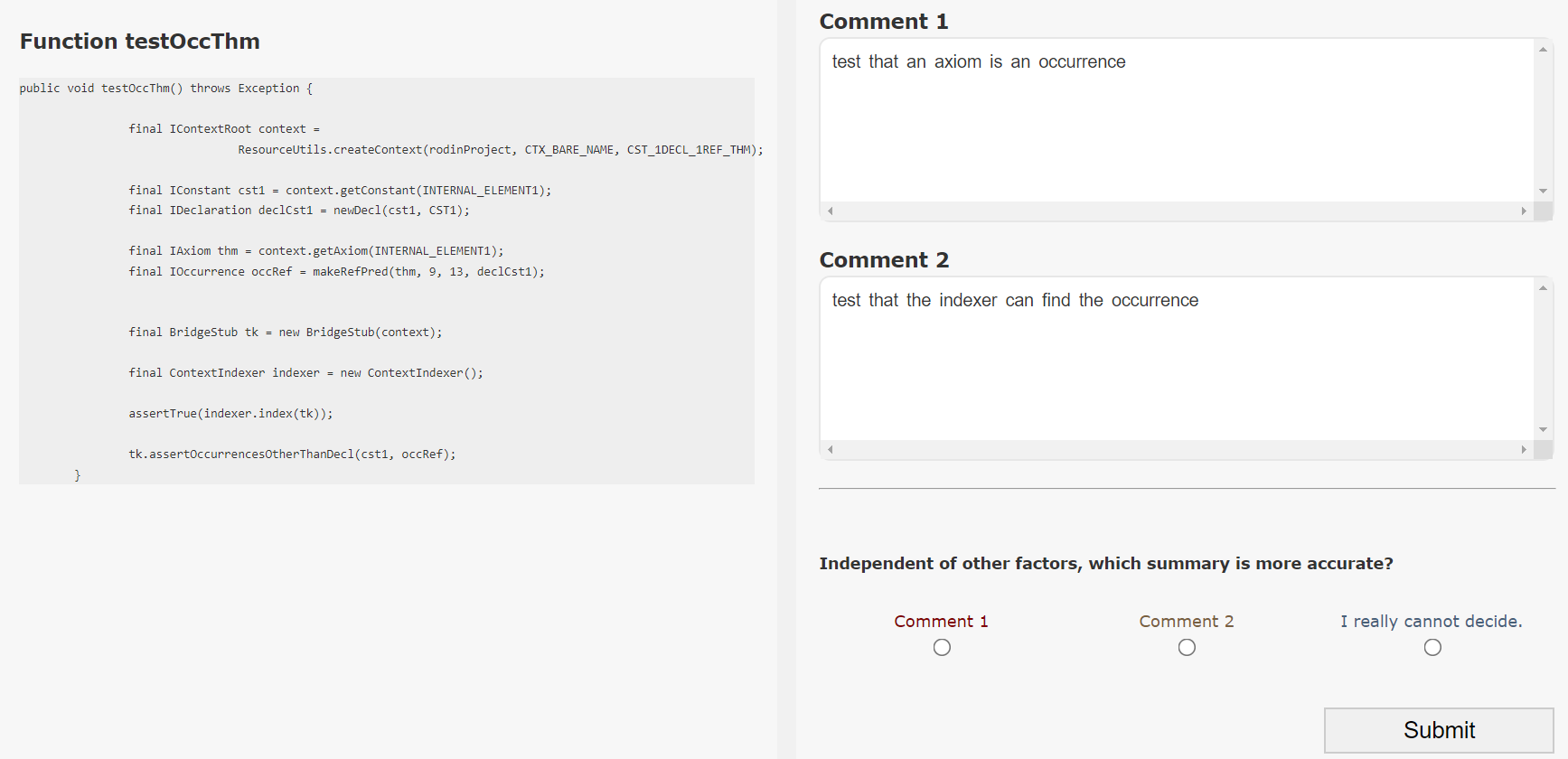}
	\caption{Screenshot of page one of the survey interface.}
	\label{fig:survey}
	\vspace{-5mm}
\end{figure}

Our study design centers around a web survey in which participants read the source code of a subroutine and a summary for that subroutine, then answer five questions (see Figure~\ref{fig:survey}).  The survey consisted of a tutorial with definitions and examples of the quality attributes, followed by four ``pages'' per subroutine.  On the first page, the people saw the subroutine source code, one summary generated by the model after fine-tuning with CCE, one summary generated by the model after fine-tuning with { \scriptsize \texttt{use-seq}} (order of summaries in UI was random and did not indicate its origin to avoid demand characteristic bias~\cite{dell2012yours}), and the following question:

\begin{description}
	\item[Q1] Independent of other factors, which summary is more accurate?
\end{description}

The next page showed the same information, but with the following two questions.  We asked these questions in a separate page to avoid a bias from showing a positively-worded question alongside a negatively-worded one~\cite{chyung2018evidence}:

\begin{description}
	\item[Q2] Which summary is missing more information that is important for understanding the method?
	\item[Q3] Which summary contains more unnecessary information?
\end{description}

The next page showed the same information, but with the following question:

\begin{description}
	\item[Q4] Overall, which summary is better in your opinion?
\end{description}

And finally, the last page showed the same information, but with a reference summary provided on the left side of the screen and the following question:

\begin{description}
	\item[Q5] Which summary is more similar to this third summary on the left?
\end{description}

It is prohibitively expensive to run a human study with all models and datasets from our experiment, so we focus on comparing CCE to { \scriptsize \texttt{use-seq}} in the following conditions:

\textbf{Model} We use the summaries generated by the LLaMA 7B model.  We focus on this model because it had the highest performance in our experiment in the previous two sections, and so is likely the model closest what could eventually be used in production.

\textbf{Subroutines} We used 56 Java methods in our study.  We sourced these by randomly sampling 60 methods from the test set of the { \texttt{funcom-java-long}} dataset in our experiment in the previous sections, then removing samples where CCE to {  \scriptsize\texttt{use-seq}} led to the model generating the identical summary.  The sample size of 50-60 is suitable because it is large enough to provide a meaningfully representative sample, but small enough to be evaluated within 90 minutes, after which fatigue bias may become a major factor~\cite{sievertsen2016cognitive}.

\textbf{Participants}  We used the Prolific\footnote{https://www.prolific.co/} platform to recruit 29 English-speaking participants who were at least 25 years of age, hold a degree in Computer Science or Engineering, and self-report experience with Java.

%% file: humanstudyresult.tex
\section{Human Study Results}

\begin{figure}[!h]
    \centering
	\includegraphics[width=12cm]{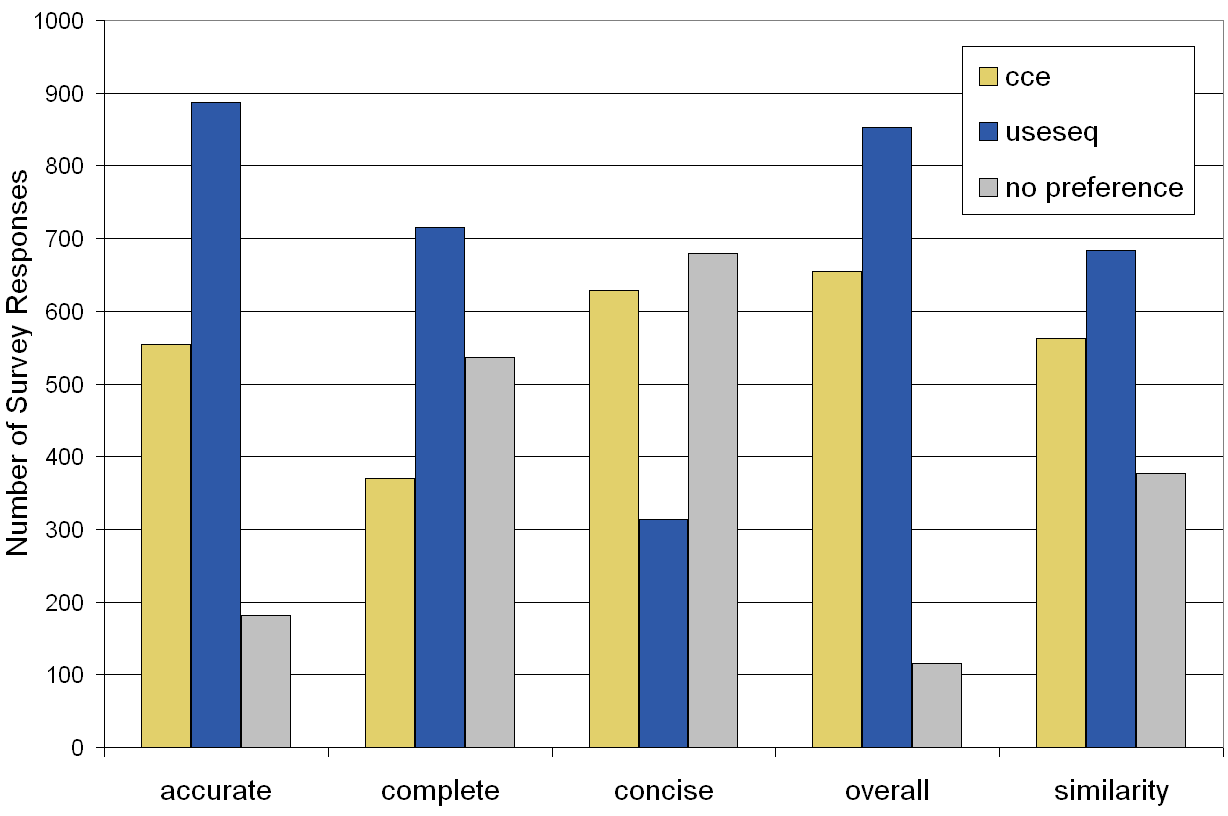}
	\caption{Aggregate survey responses.  Each column indicates the total number of survey replies for a quality attribute and summaries from one loss function.  With 29 participants and 56 functions, there are 1624 responses per attribute.  For example, there were $\sim$900 responses in which participants preferred {\scriptsize \texttt{use-seq}} in terms of accuracy versus $\sim$550 for CCE.}
	\label{fig:aggregatesurvey}
    \vspace{-4mm}
\end{figure}

This section discusses the results of our human study.  Figure~\ref{fig:aggregatesurvey} shows the number of responses from all users for each quality attribute.  The way to read this figure is that out of 1624 total survey responses for each question (29 participants x 56 methods), e.g., 850 preferred {\scriptsize \texttt{use-seq}} overall versus 650 for CCE (52\% vs 40\%, with the balance undecided).  The human raters had a stronger preference for {\scriptsize \texttt{use-seq}} than CCE in overall subjective judgment.

One reason appears to be higher levels of accuracy for {\scriptsize \texttt{use-seq}}.  The strongest difference is evident for Q1 about accuracy, where over 55\% of responses favored {\scriptsize \texttt{use-seq}} and 34\% favored CCE, with only around 11\% unable to decide.  The accuracy responses seem associated with completeness, where {\scriptsize \texttt{use-seq}} also performs well.  This result aligns with our experimental results in Section~\ref{sec:experimentresults}, where {\scriptsize \texttt{use-seq}} tends to find more words to explain what the components of the subroutine do, such as using the word ``control'' to modify ``fps'' in Example 1, Table~\ref{tab:examples}.  

The {\scriptsize \texttt{use-seq}} approach does not always perform well in completeness, as Example 2, Table~\ref{tab:examples} shows.  But in general, the errors {\scriptsize \texttt{use-seq}} makes tend to be in generating summaries that are too verbose.  CCE outperforms {\scriptsize \texttt{use-seq}} by a 2-1 margin in conciseness.  In addition, similarity to the reference does not seem to be associated with higher overall preference.  The users did tend to view {\scriptsize \texttt{use-seq}} as generating summaries more similar to the reference than CCE, though the difference is the lowest of any quality attribute, and the users did not see the reference until the last question page.

\begin{figure}[!h]
	\centering

	\vspace{-2mm}

	\includegraphics[height=6cm]{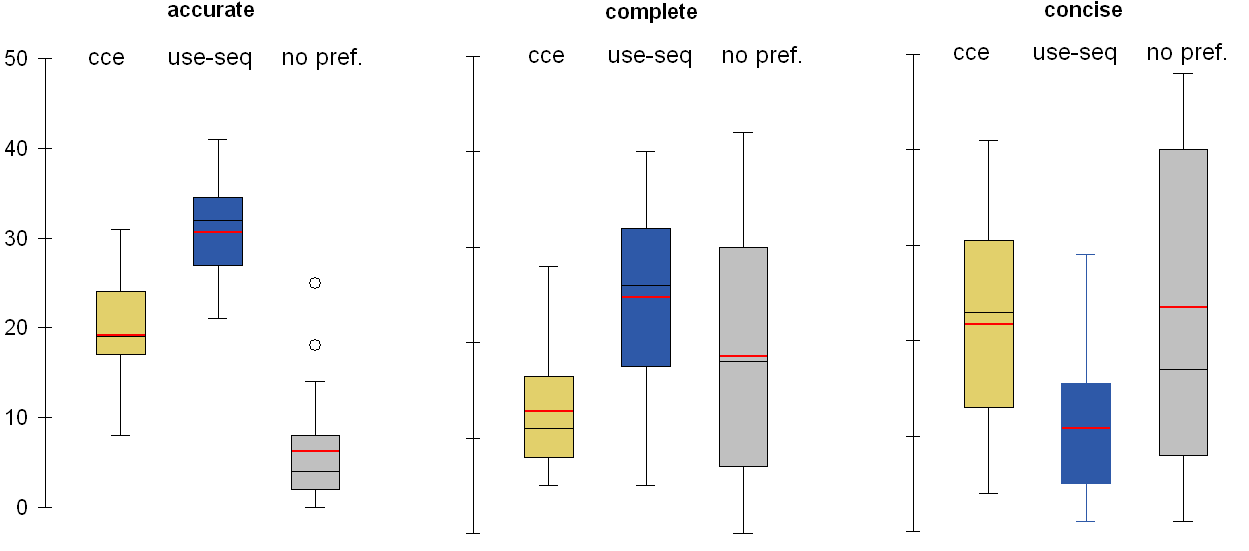}
	
	\vspace{4mm}
	
	\includegraphics[height=6.2cm]{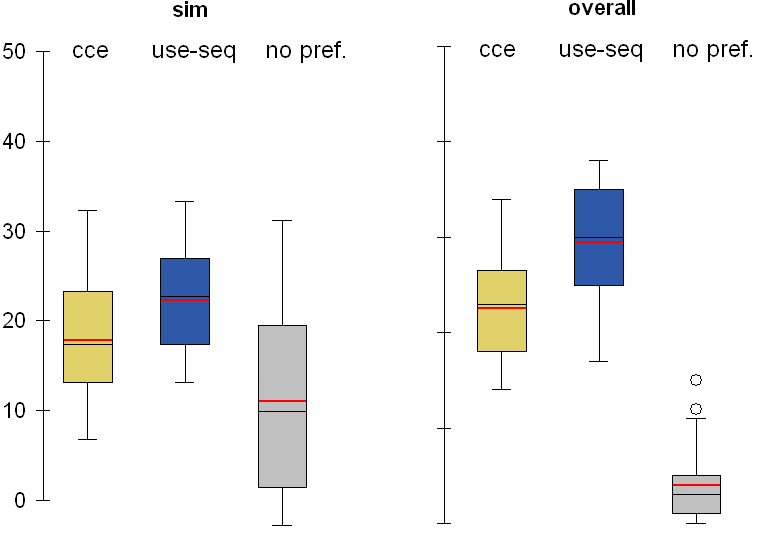}
	
	\vspace{4mm}

	\small
	\begin{tabular}{lllll|lll}
		& min & max & med & mean  & Qobs  & Qcrit & p               \\ \cline{2-8} 
		\multicolumn{1}{l|}{accurate}   & 21  & 41  & 32  & 30.58 & 41.45 & 5.99  & \textless{}0.01 \\
		\multicolumn{1}{l|}{complete}   & 5   & 40  & 26  & 24.69 & 15.52 & 5.99  & \textless{}0.01 \\
		\multicolumn{1}{l|}{concise}    & 1   & 29  & 8   & 10.83 & 13.50 & 5.99  & \textless{}0.01 \\
		\multicolumn{1}{l|}{overall}    & 17  & 38  & 30  & 29.48 & 46.41 & 5.99  & \textless{}0.01 \\
		\multicolumn{1}{l|}{similarity} & 9   & 34  & 24  & 23.59 & 7.122 & 5.99  & \textless{}0.01
	\end{tabular}

    \vspace{2mm}

	\caption{Table shows a statistical summary of responses for {\scriptsize \texttt{use-seq}} grouped by participant, followed by Friedman significance test results.  For example, an ``average'' participant, in terms of accuracy, preferred 30.58 summaries by {\scriptsize \texttt{use-seq}} compared to $\sim$19 for CCE, and had no preference for $\sim$6.  Boxplots show comparisons grouped by participant for all questions.}
	\label{fig:boxplots}
\end{figure}

To provide another view of the data, we also grouped the responses by participant (Figure~\ref{fig:boxplots}).  We define ``group by participant'' as the number of times that participant responded with each option.  The mean value is visible as the red line in the boxplots, and the median value is the black middle line.  Raw minimum, maximum, median, and mean values for {\scriptsize \texttt{use-seq}} are in the table.  All participants rated all 56 methods.  An ``average'' participant rated 32 (median) or 30.58 (mean) of the methods as preferring {\scriptsize \texttt{use-seq}} in terms of accuracy, compared to fewer than 20 for CCE, and undecided for six.  

The results broadly concur with the aggregate survey responses.  The overall preference responses lean towards {\scriptsize \texttt{use-seq}}, where a typical user preferred {\scriptsize \texttt{use-seq}} in 30 (median) or 29.48 (mean) of 56 methods.  The accuracy, completeness, and overall similarity likewise lean towards {\scriptsize \texttt{use-seq}}, though again conciseness does not, and again similarity only slightly favors {\scriptsize \texttt{use-seq}}.

As an additional check, we performed a Friedman paired statistical test for each of the quality attributes, grouped by participant.  We report $Q_{observed}$, $Q_{critical}$, and $p$ values for these tests in Figure~\ref{fig:boxplots}.  The Friedman test is the correct test because there are three sets of values, and each set of values are paired because they are tied to a specific participant over the same set of methods.  It is also a non-parametric test, making it suitable for use in preference rankings where preference itself is measurable, but the degree of difference in preference is difficult to measure~\cite{sheldon1996use}.  In general, strongest statistical significance occurred for the overall and accuracy questions, which supports the general conclusion that participants preferred summaries from {\scriptsize \texttt{use-seq}} in part due to the higher overall accuracy.

\subsection{Threats to Validity}
We also divide the key threats to validity in this study into three different categories as in the previous section. The internal threat to validity involves our choice for the model, the subroutines, and the participants. We attempted to mitigate these risks by choosing a random sample of subroutines and human programmers large enough to provide a representative sample, but the risk remains that different results are possible with different inputs. Another internal threat to validity is that the participants might fake the information in the online platforms such as Prolific or the participants might just click through without reading the information carefully. We mitigate this threat by manually inspecting the time that the participants spent on each question. We do not include the programming knowledge test because those questions can be easily answered by using AI-based tools such as ChatGPT~\cite{ghorbani2023autonomy}. The construct threats to validity include the survey interface design and the wording of the questions. We attempted to mitigate threats from the survey interface and questions by adhering to practice in related work, it is possible that differently-worded questions or interface design could change the study results. The external threat to validity is that the programmer that we hired to evaluate the summary might not be good at Java programming. We mitigated this threat by setting the constraint that the participants should be at least 25 years old, which eliminates the undergraduate students.

%% file: conclusion.tex
\section{Discussion}
\label{sec:discussions}
The integration of semantics similarity to the loss function is not limited to USE score. The use of the USE scores demonstrated that it is possible to integrate the semantic similarity into the loss function to give the reward for the correct prediction and penalize the wrong prediction. The future experiments include exploring the use of different semantic metrics between reference summary and predicted summary such as CrystalBLEU~\cite{eghbali2023crystalbleU} and the use of semantic metrics between source code and predicted summary such as SIDE~\cite{mastropaolo2023evaluating}.

Our approach is likely to have an impact beyond code summarization.  Within the field of Software Engineering, several tasks involve learning representations of code and writing or modifying text, such as bug report description and triage, test generation, and automatic privacy notification.  In fields beyond SE, {\scriptsize \texttt{use-seq}} may have applications in numerous areas of text generation, such as neural machine translation and general tasks involving fine-tuning of LLMs.  We demonstrated how our approach can be used in fine-tuning LLMs for one task, with other areas as future work.

\section{Conclusion}
\label{sec:conclusion}

This paper moves the state-of-the-art forward in five ways:
\begin{enumerate}
	\item We introduce a procedure for using a semantic similarity loss function (that we call {\scriptsize \texttt{use-seq}}), that we designed for the software engineering task of code summarization.
	\vspace{0.5mm}
	\item We evaluate our approach with four purpose-built code summarization models over datasets in two programming languages against three baselines (CCE, BLEU, SimiLE).  We show how {\scriptsize \texttt{use-seq}} achieves improvements in several conditions.
	\item We propose and implement code summarization as fine-tuning of two industrial language models. 
	\vspace{0.5mm}
	\item We evaluate our approach over a Java dataset against the state-of-the-practice CCE using automated metrics.
	\vspace{0.5mm}
	\item We perform a study with 29 human programmers to evaluate summaries for 54 Java methods.  We compare and contrast summaries generated by LLaMA with {\scriptsize \texttt{use-seq}} versus LLaMA with CCE. 
\end{enumerate}

Overall, we found that {\scriptsize \texttt{use-seq}} improves purpose-built code summarization approaches by 2-3\% when measured by automated metrics METEOR and USE, and up to 12\% when measured by BLEU.  This is a strongly positive result considering that the improvement 1) is consistent over multiple approaches and datasets, and 2) comes at practically no cost or additional training procedure complexity.  The {\scriptsize \texttt{use-seq}} loss function may be used as a drop-in replacement for CCE, unlike baselines which require additional steps.

We have designed an evaluated our approach for the problem of code summarization in the domain of software engineering.  Key considerations we made for our target problem include 1) computing USE to compare summaries, which is the semantic similarity measure that recent literature~\cite{haque2022semantic} found is most associated with human programmer judgments of summary similarity, 2) we mask semantic similarities to avoid inappropriate penalties for correct word predictions when the sequence similarity is overall poor (and avoid rewarding incorrect words when the sequence similarity is good), and 3) we do \emph{not} include a length penalty as baselines from NLP do, since people tend to prefer accuracy to conciseness.

To encourage maximum reproducibility and accessibility of our research, we provide our implementation code, training data, evaluation scripts, and further results via an online repository in Code Availability and Data Availability Section. Note that the full code of the examples provided in Section~\ref{sec:experimentresults} of this paper are available by using the ID number under the example method name with the \emph{fid} index in the {\scriptsize \texttt{funcom-java-long}} dataset we provide in Data Availability Section.

%% file: wileyNJDv5_AMA.bbl
\begin{thebibliography}{10}
\providecommand \doibase [0]{http://dx.doi.org/}%

\bibitem{haiduc2010use}
Haiduc S, Aponte J, Moreno L, Marcus A. On the use of automated text summarization techniques for summarizing source code. In: IEEE.  2010\string:35--44

\bibitem{robillard2017demand}
Robillard MP, Marcus A, Treude C, et al. On-demand developer documentation. In: IEEE.  2017\string:479--483

\bibitem{leclair2019neural}
LeClair A, Jiang S, McMillan C. A neural model for generating natural language summaries of program subroutines. In: IEEE Press.  2019\string:795--806

\bibitem{leclair2020improved}
LeClair A, Haque S, Wu L, McMillan C. Improved code summarization via a graph neural network. In:  2020\string:184--195

\bibitem{haque2020improved}
Haque S, LeClair A, Wu L, McMillan C. Improved Automatic Summarization of Subroutines via Attention to File Context. {\it International Conference on Mining Software Repositories.} 2020.
\newblock \href {\doibase 10.1145/3379597.3387449} {doi: 10.1145/3379597.3387449}

\bibitem{tang2022ast}
Tang Z, Shen X, Li C, et al. AST-trans: code summarization with efficient tree-structured attention. In:  2022\string:150--162

\bibitem{macneil2023experiences}
MacNeil S, Tran A, Hellas A, et al. Experiences from using code explanations generated by large language models in a web software development e-book. In:  2023\string:931--937

\bibitem{ross2023programmer}
Ross SI, Martinez F, Houde S, Muller M, Weisz JD. The programmer’s assistant: Conversational interaction with a large language model for software development. In:  2023\string:491--514

\bibitem{wieting2019beyond}
Wieting J, Berg-Kirkpatrick T, Gimpel K, Neubig G. Beyond BLEU: Training Neural Machine Translation with Semantic Similarity. In:  2019\string:4344--4355

\bibitem{haque2022semantic}
Haque S, Eberhart Z, Bansal A, McMillan C. Semantic similarity metrics for evaluating source code summarization. In:  2022\string:36--47

\bibitem{touvron2023llama}
Touvron H, Lavril T, Izacard G, et al. Llama: Open and efficient foundation language models. {\it arXiv preprint arXiv:2302.13971.} 2023.

\bibitem{alpaca}
Taori R, Gulrajani I, Zhang T, et al. Stanford Alpaca: An Instruction-following LLaMA model. https://github.com/tloen/alpaca-lora;  2023.

\bibitem{alpacalora}
Wang E. Alpaca-LoRA. https://github.com/tatsu-lab/stanford\_alpaca;  2023.

\bibitem{hulora}
Hu EJ, Wallis P, Allen-Zhu Z, et al. LoRA: Low-Rank Adaptation of Large Language Models. In:  2023.

\bibitem{alon2019code2seq}
Alon U, Brody S, Levy O, Yahav E. code2seq: Generating sequences from structured representations of code. {\it International Conference on Learning Representations.} 2019.

\bibitem{alon2019code2vec}
Alon U, Zilberstein M, Levy O, Yahav E. code2vec: Learning distributed representations of code. {\it Proceedings of the ACM on Programming Languages.} 2019\string;3(POPL)\string:1--29.
\newblock \href {\doibase 10.1145/3290353} {doi: 10.1145/3290353}

\bibitem{nie2019framework}
Nie P, Rai R, Li JJ, Khurshid S, Mooney RJ, Gligoric M. A framework for writing trigger-action todo comments in executable format. In: ACM.  2019\string:385--396

\bibitem{haldar2020multi}
Haldar R, Wu L, Xiong J, Hockenmaier J. A Multi-Perspective Architecture for Semantic Code Search. {\it arXiv preprint arXiv:2005.06980.} 2020.

\bibitem{ahmad2020transformer}
Ahmad WU, Chakraborty S, Ray B, Chang KW. A Transformer-based Approach for Source Code Summarization. {\it arXiv preprint arXiv:2005.00653.} 2020.

\bibitem{feng2020codebert}
Feng Z, Guo D, Tang D, et al. CodeBERT: A Pre-Trained Model for Programming and Natural Languages. In:  2020\string:1536--1547

\bibitem{bansal2021project}
Bansal A, Haque S, McMillan C. Project-level encoding for neural source code summarization of subroutines. In: IEEE.  2021\string:253--264

\bibitem{zugner2021languageagnostic}
Z{\"u}gner D, Kirschstein T, Catasta M, Leskovec J, G{\"u}nnemann S. Language-Agnostic Representation Learning of Source Code from Structure and Context. In:  2021.

\bibitem{liu2021retrievalaugmented}
Liu S, Chen Y, Xie X, Siow JK, Liu Y. Retrieval-Augmented Generation for Code Summarization via Hybrid {\{}GNN{\}}. In:  2021.

\bibitem{mastropaolo2021studying}
Mastropaolo A, Scalabrino S, Cooper N, et al. Studying the usage of text-to-text transfer transformer to support code-related tasks. In: IEEE.  2021\string:336--347.

\bibitem{kuang2022code}
Kuang L, Zhou C, Yang X. Code comment generation based on graph neural network enhanced transformer model for code understanding in open-source software ecosystems. {\it Automated Software Engineering.} 2022\string;29(2)\string:43.
\newblock \href {\doibase 10.1007/s10515-022-00341-1} {doi: 10.1007/s10515-022-00341-1}

\bibitem{li2022setransformer}
Li Z, Wu Y, Peng B, et al. Setransformer: A transformer-based code semantic parser for code comment generation. {\it IEEE Transactions on Reliability.} 2022.
\newblock \href {\doibase 10.1109/TR.2022.3154773} {doi: 10.1109/TR.2022.3154773}

\bibitem{khan2022automatic}
Khan JY, Uddin G. Automatic code documentation generation using gpt-3. In:  2022\string:1--6.

\bibitem{ahmed2022few}
Ahmed T, Devanbu P. Few-shot training LLMs for project-specific code-summarization. In:  2022\string:1--5.

\bibitem{gu2022assemble}
Gu J, Salza P, Gall HC. Assemble foundation models for automatic code summarization. In: IEEE.  2022\string:935--946.

\bibitem{10.1145/3611643.3613090}
Su CY, Bansal A, Jain V, Ghanavati S, McMillan C. A Language Model of Java Methods with Train/Test Deduplication. In: ESEC/FSE 2023. Association for Computing Machinery 2023; New York, NY, USA\string:2152–2156

\bibitem{gao2023code}
Gao S, Gao C, He Y, et al. Code Structure--Guided Transformer for Source Code Summarization. {\it ACM Transactions on Software Engineering and Methodology.} 2023\string;32(1)\string:1--32.

\bibitem{geng2023interpretation}
Geng M, Wang S, Dong D, et al. Interpretation-based Code Summarization. In:  2023.

\bibitem{zhang2023ga}
Zhang M, Zhou G, Yu W, Huang N, Liu W. Ga-scs: Graph-augmented source code summarization. {\it ACM Transactions on Asian and Low-Resource Language Information Processing.} 2023\string;22(2)\string:1--19.

\bibitem{gao2023encosum}
Gao Y, Zhang H, Lyu C. EnCoSum: enhanced semantic features for multi-scale multi-modal source code summarization. {\it Empirical Software Engineering.} 2023\string;28(5)\string:126.

\bibitem{wang2023intra}
Wang Z, Yu X, Feng Y, Zhao D. An Intra-Class Relation Guided Approach for Code Comment Generation. In:  2023\string:1291--1303.

\bibitem{geng2024large}
Geng M, Wang S, Dong D, et al. Large Language Models are Few-Shot Summarizers: Multi-Intent Comment Generation via In-Context Learning.  2024.

\bibitem{song2019survey}
Song X, Sun H, Wang X, Yan J. A Survey of Automatic Generation of Source Code Comments: Algorithms and Techniques. {\it IEEE Access.} 2019.

\bibitem{jin2023binary}
Jin X, Larson J, Yang W, Lin Z. Binary code summarization: Benchmarking chatgpt/gpt-4 and other large language models. {\it arXiv preprint arXiv:2312.09601.} 2023.

\bibitem{sun2023automatic}
Sun W, Fang C, You Y, et al. Automatic Code Summarization via ChatGPT: How Far Are We?. {\it arXiv preprint arXiv:2305.12865.} 2023.

\bibitem{wang2021codet5}
Wang Y, Wang W, Joty S, Hoi SC. {C}ode{T}5: Identifier-aware Unified Pre-trained Encoder-Decoder Models for Code Understanding and Generation. In:  Moens MF, Huang X, Specia L, Yih SWt. \kern-2pt, eds. {\it Proceedings of the 2021 Conference on Empirical Methods in Natural Language Processing}Association for Computational Linguistics 2021; Online and Punta Cana, Dominican Republic\string:8696--8708

\bibitem{bender2021danger}
Bender EM, Gebru T, McMillan-Major A, Shmitchell S. On the Dangers of Stochastic Parrots: Can Language Models Be Too Big?. In: FAccT '21. Association for Computing Machinery 2021; New York, NY, USA\string:610–623

\bibitem{papineni2002bleu}
Papineni K, Roukos S, Ward T, Zhu WJ. BLEU: a method for automatic evaluation of machine translation. In: Association for Computational Linguistics.  2002\string:311--318

\bibitem{banerjee2005meteor}
Banerjee S, Lavie A. METEOR: An automatic metric for MT evaluation with improved correlation with human judgments. In:  2005\string:65--72.

\bibitem{eghbali2023crystalbleU}
Eghbali A, Pradel M. CrystalBLEU: Precisely and Efficiently Measuring the Similarity of Code. In: ASE '22. Association for Computing Machinery 2023; New York, NY, USA

\bibitem{mastropaolo2023evaluating}
Mastropaolo A, Ciniselli M, Di~Penta M, Bavota G. Evaluating Code Summarization Techniques: A New Metric and an Empirical Characterization. In: ICSE '24. Association for Computing Machinery 2024; New York, NY, USA

\bibitem{ranzato2016sequence}
Ranzato M, Chopra S, Auli M, Zaremba W. Sequence level training with recurrent neural networks. In:  2016.

\bibitem{pasunuru2018multi}
Pasunuru R, Bansal M. Multi-Reward Reinforced Summarization with Saliency and Entailment. In:  2018\string:646--653

\bibitem{chen2020deep}
Chen Y, Lu X. Deep category-level and regularized hashing with global semantic similarity learning. {\it IEEE transactions on cybernetics.} 2020\string;51(12)\string:6240--6252.
\newblock \href {\doibase 10.1109/TCYB.2020.2964993} {doi: 10.1109/TCYB.2020.2964993}

\bibitem{nakatani2022comparing}
Nakatani Y, Kajiwara T, Ninomiya T. Comparing BERT-based Reward Functions for Deep Reinforcement Learning in Machine Translation. In:  2022\string:37--43.

\bibitem{yasui2019using}
Yasui G, Tsuruoka Y, Nagata M. Using semantic similarity as reward for reinforcement learning in sentence generation. In:  2019\string:400--406

\bibitem{cer2018universal}
Cer D, Yang Y, Kong Sy, et al. Universal sentence encoder. {\it arXiv preprint arXiv:1803.11175.} 2018.

\bibitem{korbak2023pretraining}
Korbak T, Shi K, Chen A, et al. Pretraining language models with human preferences. {\it arXiv preprint arXiv:2302.08582.} 2023.

\bibitem{wohlin2012experimentation}
Wohlin C, Runeson P, H{\"o}st M, Ohlsson MC, Regnell B, Wessl{\'e}n A. {\it Experimentation in software engineering}.
\newblock Springer Science \& Business Media, 2012.

\bibitem{hellendoorn2021growing}
Hellendoorn VJ, Sawant AA. The growing cost of deep learning for source code. {\it Communications of the ACM.} 2021\string;65(1)\string:31--33.
\newblock \href {\doibase 10.1145/3501261} {doi: 10.1145/3501261}

\bibitem{xu2022systematic}
Xu FF, Alon U, Neubig G, Hellendoorn VJ. A systematic evaluation of large language models of code. In:  2022\string:1--10

\bibitem{wieting2018paranmt}
Wieting J, Gimpel K. ParaNMT-50M: Pushing the Limits of Paraphrastic Sentence Embeddings with Millions of Machine Translations. In:  2018\string:451--462

\bibitem{wu2016google}
Wu Y, Schuster M, Chen Z, et al. Google's neural machine translation system: Bridging the gap between human and machine translation. {\it arXiv preprint arXiv:1609.08144.} 2016.

\bibitem{leclair2019recommendations}
LeClair A, McMillan C. Recommendations for Datasets for Source Code Summarization. In:  2019\string:3931--3937

\bibitem{allamanis2019adverse}
Allamanis M. The adverse effects of code duplication in machine learning models of code. In:  2019\string:143--153

\bibitem{shi2022are}
Shi L, Mu F, Chen X, et al. Are We Building on the Rock? On the Importance of Data Preprocessing for Code Summarization. In: ESEC/FSE 2022. Association for Computing Machinery 2022\string:107–119

\bibitem{haque2021action}
Haque S, Bansal A, Wu L, McMillan C. Action Word Prediction for Neural Source Code Summarization. {\it 28th IEEE International Conference on Software Analysis, Evolution and Reengineering.} 2021.
\newblock \href {\doibase 10.1109/SANER50967.2021.00038} {doi: 10.1109/SANER50967.2021.00038}

\bibitem{roy2021reassessing}
Roy D, Fakhoury S, Arnaoudova V. Reassessing Automatic Evaluation Metrics for Code Summarization Tasks. In:  2021

\bibitem{radford2019language}
Radford A, Wu J, Child R, et al. Language models are unsupervised multitask learners. {\it OpenAI blog.} 2019\string;1(8)\string:9.

\bibitem{nanogpt}
Karpathy A. nanoGPT: The simplest, fastest repository for training/finetuning medium-sized GPTs.. https://github.com/karpathy/nanoGPT;  2023.

\bibitem{su2024distilled}
Su CY, McMillan C. Distilled GPT for source code summarization. {\it Automated Software Engineering.} 2024\string;31(1)\string:22.

\bibitem{sridhara2010towards}
Sridhara G, Hill E, Muppaneni D, Pollock L, Vijay-Shanker K. Towards automatically generating summary comments for java methods. In: ACM.  2010\string:43--52

\bibitem{ferretti2023naturalness}
Ferretti C, Saletta M. Naturalness in Source Code Summarization. How Significant is it?. In: IEEE.  2023\string:125--134.

\bibitem{rani2023decade}
Rani P, Blasi A, Stulova N, Panichella S, Gorla A, Nierstrasz O. A decade of code comment quality assessment: A systematic literature review. {\it Journal of Systems and Software.} 2023\string;195\string:111515.

\bibitem{dell2012yours}
Dell N, Vaidyanathan V, Medhi I, Cutrell E, Thies W. " Yours is better!" participant response bias in HCI. In:  2012\string:1321--1330

\bibitem{chyung2018evidence}
Chyung SY, Barkin JR, Shamsy JA. Evidence-based survey design: The use of negatively worded items in surveys. {\it Performance Improvement.} 2018\string;57(3)\string:16--25.
\newblock \href {\doibase 10.1002/pfi.21749} {doi: 10.1002/pfi.21749}

\bibitem{sievertsen2016cognitive}
Sievertsen HH, Gino F, Piovesan M. Cognitive fatigue influences students’ performance on standardized tests. {\it Proceedings of the National Academy of Sciences.} 2016\string;113(10)\string:2621--2624.
\newblock \href {\doibase 10.1073/pnas.1516947113} {doi: 10.1073/pnas.1516947113}

\bibitem{sheldon1996use}
Sheldon MR, Fillyaw MJ, Thompson WD. The use and interpretation of the Friedman test in the analysis of ordinal-scale data in repeated measures designs. {\it Physiotherapy Research International.} 1996\string;1(4)\string:221--228.
\newblock \href {\doibase 10.1002/pri.66} {doi: 10.1002/pri.66}

\bibitem{ghorbani2023autonomy}
Ghorbani A, Cassee N, Robinson D, et al. Autonomy Is an Acquired Taste: Exploring Developer Preferences for GitHub Bots. In: ICSE '23. IEEE Press 2023\string:1405–1417

\end{thebibliography}
